\documentclass[]{pasj01}
\draft
\usepackage{lscape}

\begin{document} 
\Received{}
\Accepted{}

\title{Properties of Molecular Gas in Galaxies in the Early and Mid-Stage of Interaction. II. Molecular Gas Fraction}

\author{Hiroyuki \textsc{Kaneko},\altaffilmark{1,2}
	Nario \textsc{Kuno},\altaffilmark{3}
	Daisuke \textsc{Iono},\altaffilmark{2,4}
	Yoichi \textsc{Tamura},\altaffilmark{5}
	Tomoka \textsc{Tosaki},\altaffilmark{6}
	Kouichiro \textsc{Nakanishi},\altaffilmark{2,4}
  	and Tsuyoshi \textsc{Sawada}\altaffilmark{2,7}
	}
\email{kaneko.hiroyuki@nao.ac.jp}
\altaffiltext{1}{Nobeyama Radio Observatory, 462-2, Minamimaki, Minamisaku, Nagano, 384-1305, Japan}
\altaffiltext{2}{National Astronomical Observatory of Japan,  2-21-1 Osawa, Mitaka, Tokyo, 181-8588, Japan}
\altaffiltext{3}{Graduate School of Pure and Applied Sciences, University of Tsukuba, 1-1-1 Tennodai, Tsukuba, Ibaraki, 305-8577, Japan}
\altaffiltext{4}{Department of Astronomical Science, The Graduate University for Advanced Studies, 2-21-1 Osawa, Mitaka, Tokyo, 181-8588, Japan}
\altaffiltext{5}{Institute of Astronomy, The University of Tokyo, 2-21-1, Osawa, Mitaka, Tokyo 181-0015}
\altaffiltext{6}{Department of Geoscience, Joetsu University of Education, Joetsu, Niigata 943-8512}
\altaffiltext{7}{Joint ALMA Observatory, Alonso de Cordova 3107, Vitacura, Santiago, Chile}


\maketitle

\begin{abstract}
We have investigated properties of the interstellar medium in interacting galaxies in early and mid-stage using mapping data of $^{12}$CO(\textit{J}~=~1--0) and H\emissiontype{I}.
Assuming the standard CO-H$_{2}$ conversion factor, we found no difference in molecular gas mass, atomic gas mass, and total gas mass (a sum of atomic and molecular gas mass)  between interacting galaxies and isolated galaxies.
However, interacting galaxies have a higher global molecular gas fraction $f_{\rm{mol}}^{\rm{global}}$, a ratio of molecular gas mass to total gas mass averaged over a whole galaxy, (0.71 $\pm$ 0.15) than isolated galaxies (0.52 $\pm$ 0.18).
The distribution of local molecular gas fraction $f_{\rm{mol}}$, a ratio of the surface density of molecular gas to that of total gas, is different from the distribution in typical isolated galaxies.
By a pixel-to-pixel comparison, isolated spiral galaxies show a gradual increase in $f_{\rm{mol}}$ along the surface density of total gas until it is saturated at 1.0, while interacting galaxies show no clear relation.
We performed pixel-to-pixel theoretical model fittings varying metallicity and external pressure.
According to the model fitting, external pressure can explain the trend of  $f_{\rm{mol}}$ in the interacting galaxies.
Assuming a half of the standard CO-H$_{2}$ conversion factor for interacting galaxies, the results of pixel-to-pixel theoretical model fitting get worse than adopting the standard conversion factor, although $f_{\rm{mol}}^{\rm{global}}$ of interacting galaxies (0.62 $\pm$ 0.17) becomes the same as in isolated galaxies.
We conclude that external pressure occurs due to the shock prevailing over a whole galaxy or due to collisions between giant molecular clouds even in the early stage of the interaction.
The external pressure accelerates an efficient transition from atomic gas to molecular gas.
Regarding chemical time-scale, high $f_{\rm{mol}}$ can be achieved at the very early stage of interaction even if shock induced by the collision of galaxies ionises interstellar gas.
\end{abstract}

\section{Introduction}
The morphological and chemical evolutions of galaxies are dominated by interstellar medium (ISM) through star-formation activity. 
While the ISM works as a fuel for the birth of stars, heavy elements made in stars spread out into interstellar space through supernovae and AGB stars at the end phase of stars.
The supernovae may trigger star formation by compressing the surrounding ISM.
Repetition of the birth and death of stars changes the galaxy-scale environments.
It is known that the early-type spiral galaxies are known to have high molecular-to-atomic gas ratios \citep{OR09,Boselli97,Sauty03,YK89}.
A global molecular gas fraction, which is the same as molecular-to-atomic gas ratio practically, also varies even within a galaxy \citep{Honma95,Tosaki11}.
\citet{Tanaka14} showed that radial distributions of a local molecular gas fraction of nearby spiral galaxies can be explained by theoretical model fitting based on \citet{Elmegreen93} using the combination of observational data.
In a galaxy cluster, H\emissiontype{I} is deficient \citep{GH83}, while CO gas is not deficient \citep{KY86, KY89, Stark86, Nakanishi06}.
These studies show that ram pressure effectively and selectively removes H\emissiontype{I}, which typically resides in an outer region farther than molecular gas, from a galaxy into intergalactic space.
The properties of ISM are highly connected to the environments of galaxies.

A galaxy interaction is one of the biggest events for galaxies.
Dynamical and morphological changes happen not only on stars but also to the ISM through the collision of galaxies.
It is also confirmed that interacting galaxies show violent star-formation activity (e.g., \cite{Bushouse87}). 
Most (ultra-)luminous infrared galaxies (U/LIRGs), which show active star-formation, are merging galaxies (a late stage of the interaction) (e.g., \cite{Borne00}).
Regarding these facts, properties of ISM in interacting galaxies should be affected during the interaction.
It is said that molecular gas mass is enhanced by the interaction \citep{Casasola04,Combes94}.
\citet{MS89} showed a ratio of molecular gas to atomic gas is higher in interacting galaxies than that in isolated galaxies on a galaxy scale.
These results, however, are derived from the CO observations aiming at the centre of galaxies, which may underestimate molecular gas mass and have no information on the distribution of molecular gas.
What is more, most investigated interacting galaxies are in the late stage of the interaction showing highly enhanced star-formation activity because they are thought to harbour plenty of molecular gas.
Properties of the ISM of interacting galaxies in the early and mid-stage are important since the ISM should fulfil the conditions of active star formation before the starburst.
However, these properties have not been well investigated.
This is because weaker star-formation activity than interacting galaxies in the late stage (e.g., U/LIRGs) implies weaker emission from molecular gas.
Understanding properties of the ISM in the early and mid-stage of the interaction will help us discover the origin of the burst of star formation in the late stage of the interaction such as U/LIRGs.

\citet{Kaneko13} (Paper I henceforth) showed $^{12}$CO(\textit{J} = 1--0) (hereafter CO) mapping data of four interacting galaxies in the early and mid-stage of the interaction.
These mapping data are suitable for investigations of properties of the ISM undergoing the effects of the interaction.
In this paper, we studied properties of the ISM in interacting galaxies in the early and mid-stage of the interaction in the context of gas content and the molecular gas fraction.
The molecular gas fraction is examined at both the galaxy scale and kpc scale, which can resolve galactic structures.
Star-formation activity in interacting galaxies in the early and mid-stage of the interaction will be discussed in a forthcoming paper.

\section{Data}
In order to investigate properties of the ISM in interacting galaxies, we need data that cover a whole system of interacting galaxies and resolve galactic structures.
Here, we use a data set of tracers of diffuse molecular gas (CO), atomic gas (H\emissiontype{I}), old stars (\textit{K$_{s}$}), and star formation  (H$\alpha$, FUV, 8 $\mu$m, and 24 $\mu$m).
For comparison, we also use the data set of CO, H\emissiontype{I}, and \textit{K$_{s}$}-band for isolated galaxies.

\subsection{CO data}
For a sample of interacting galaxies, we used the CO data obtained with the 45-m telescope at Nobeyama Radio Observatory (NRO) of four interacting galaxies, Arp 84, VV 219, VV 254, and the Antennae Galaxies from Paper I.
In this paper, the definition of ``the bridge region'' of VV 254 is based on \citet{Braine03}.
We assume that the inclination of the bridge region is estimated to be \timeform{67D}, which is an average of the inclination of two constituent galaxies, UGC 12914 (\timeform{61D}) and UGC 12915 (\timeform{73D}). 
Details are summarised in Paper I. 

For comparison, a control sample of isolated spiral galaxies was obtained from Nobeyama CO Atlas \citep{Kuno07}.
This atlas is a CO mapping survey of 40 nearby spiral galaxies using the Nobeyama 45-m telescope. 
Since the atlas covers almost the whole region of the target galaxies with the same angular resolution as our observations (\timeform{19''.3} at 115.27 GHz, which is the rest frequency of the CO line), the atlas data can be used for a direct comparison with our data of interacting galaxies.
For some galaxies in Nobeyama CO Atlas which did not obtain the data with the On-The-Fly technique\footnote{Since On-The-Fly mapping observation does not obtain the data aligning on any regular grid, the observed data should be regridded onto a regular grid with a convolution function. 
Due to Gaussian-tapered first-order Bessel function we used as a convolution function, an effective angular resolution becomes \timeform{19''.3}.} 
(i.e., the angular resolution is \timeform{16''}), we convolved them to match the angular resolution as \timeform{19''.3}.
To remove the environmental effects, we excluded galaxies that belong to the Virgo and Coma Clusters from the control sample.
The average of the distance of interacting galaxies is 38 Mpc, which is larger than that of any isolated galaxies in \citet{Kuno07}
and our interacting galaxies sample has a lower sensitivity of CO integrated intensity on average than galaxies in \citet{Kuno07} about a factor of two.
In order to have same CO sensitivity between interacting galaxies and isolated galaxies, we also remove isolated galaxies with the distance nearer than 7 Mpc.

\subsection{H\emissiontype{I} data}
For interacting galaxies, we used the H\emissiontype{I} data obtained with VLA by \citet{Iono05}, except for the Antennae Galaxies whose data was taken from \citet{Hibbard01} using VLA.

The H\emissiontype{I} data were taken from THINGS\footnote{This work made use of THINGS, `The H\emissiontype{I} Nearby Galaxy Survey' \citep{Walter08}} for six isolated galaxies.
We also gathered the H\emissiontype{I} data from the literature for isolated galaxies which surveyed by the Nobeyama CO Atlas but were not observed by THINGS.
The final control sample of isolated galaxies comprises 13 galaxies.

\subsection{\textit{K$_{s}$} data}
In order to normalise the difference of size of each galaxy with stellar mass, we utilised the \textit{K$_{s}$} data.
All data were taken from the 2MASS catalogue\footnote{This publication makes use of data products from the Two Micron All Sky Survey, which is a joint project of the University of Massachusetts and the Infrared Processing and Analysis Center/California Institute of Technology, funded by the National Aeronautics and Space Administration and the National Science Foundation} \citep{Skrutskie06, Jarrett03}.

\subsection{Star formation tracers}
H$\alpha$ data are collected from published papers.
Although H$\alpha$ images for the progenitors of Arp 84 (NGC 5394 and NGC 5395) and VV 254 (UGC 12914 and UGC 12915) are available \citep{Bushouse87, Epinat08}, field-of-view for these H$\alpha$ images is limited.
For this reason, we used GALEX FUV images instead when we need to make maps for Arp 84 and VV 254.

Both H$\alpha$ and FUV emitted from OB stars are markedly absorbed by surrounding dust.
Therefore, the star-formation rate (hereafter SFR) derived from these data should be corrected.
For this reason, we also made use of Spitzer MIPS 24 $\mu$m data, which is thought to be emitted from the dust.
Since MIPS 24 $\mu$m image of Arp 84 is saturated, we used IRAC 8 $\mu$m data.\\
\begin{table*}
	\tbl{Control isolated galaxies}{
	\begin{tabular}{cccccc}
		\hline
		Name    &   Morphology    &   Velocity (Ref.\footnotemark[$*$])    &  Distance (Ref.\footnotemark[$*$])   & Inclination (Ref.\footnotemark[$*$])  & $R_{K20}$\\
				&                & (km s$^{-1}$) &    (Mpc)    &    (deg)  & (arcsec) \\
		(1)    &      (2)        &      (3)      &     (4)     &     (5)		& (6)\\
				\hline
				 NGC 253	 & SAB(s)c		  &       237 (1)	&     2.5(6)   &      75(1)	&	630.2\\
				 UGC 2855	& SABc			  &      1207 (1)  &    20.3 (2)  &      63 (8)	&	114.3 \\
				 NGC 3184	& SAB(rs)cd		&       594 (2)   &     8.7 (2)	 &      21 (8)	&	114.6 \\
				 NGC 3351	& SB(r)b		  &       778 (3)	&    10.1 (7) &      40 (3)	&	116.3 \\
				 NGC 3521	& SAB(rs)bc		&       792 (1)   &     7.2 (8)	 &      63 (1)	&	164.4 \\
				 NGC 3627	& SAB(s)b		 &       715 (1)   &    11.1 (9)	&      52 (1)	&	185.0  \\
				 NGC 3631	& SA(s)c		  &      1164 (2)  &    21.6 (8)	&      17 (8)	&	80.9 \\
				 NGC 4051	& SAB(rs)bc		&       725 (4)   &    17.0 (8)	&      49 (11)	&	102.6 \\
				 NGC 4102	& SAB(s)b		 &       853 (1)   &    17.0 (8)	&      56 (11)	&	68.6 \\
				 NGC 5055	& SA(rs)bc		 &       503 (1)   &     7.2 (8)	&      61 (1)	&	204.2 \\
				 NGC 5248	& SAB(rs)cd		&      1165 (1)	 &    22.7 (8)	&      40 (8)	&	111.8 \\
				 NGC 5457	& SAB(rs)cd		&       255 (5)	  &     7.2 (10)	&      18 (12)	&	236.3 \\
				 NGC 6217	& (R)SA(rs)bc  &      1355 (2)	 &    23.9 (8)	&      34 (8)	&	73.2 \\
				 NGC 6951	& SAB(rs)bc		&      1425 (1)  &    24.1 (8)	&      30 (1)	&	115.6 \\
				\hline
			\end{tabular}}
		\label{isolate}
		\begin{tabnote}
			Column (1): Galaxy name.\\
			Column (2): Morphological type taken from RC3 catalogue \citep{RC3}.\\
			Column (3): Velocity in local standard of rest.\\
			Column (4): Distance.\\
			Column (5): Inclination angle.\\
			Column (6): Radius at 20 magnitude arcsec$^{-2}$ in \textit{K$_{s}$}-band \citep{Jarrett03}\\
			\footnotemark[$*$] Reference: 1. \citet{Kuno07}; 2. \citet{Tully74}; 3. \citet{Regan01}; 4. \citet{Helfer03} \\
			5. \citet{Sofue03}; 6. \citet{Mauersberger96}; 7. \citet{Graham97}; 8. \citet{RC3}; \\
			9. \citet{Saha99}; 10. \citet{Stetson98}; 11. \citet{VS01}; 12. \citet{BGA81}
		\end{tabnote}
\end{table*}

We convolved all data with the same spatial resolution of CO, which is the lowest among the data.
A pixel size is set to a half of the original beam size.
For interacting galaxies, their linear resolutions are about 1 kpc.
A list of the control isolated galaxies is shown in table \ref{isolate}.

\section{Results}
\label{gasmass}
We examine whether an enhancement of interstellar gas mass (H\emissiontype{I}, H$_{2}$, and total gas mass) occurs in interacting galaxies by comparing with the control sample.
Molecular gas mass is derived using the following equation,
\begin{equation}
	M_\mathrm{{H_{2}}}[M_{\odot}] = 3.9 \times 10^{-17} X_{\mathrm{CO}} D^{2} S_{\mathrm{CO}}, 
	\label{H2}
\end{equation}
where $X_{\mathrm {CO}}$ is the CO-H$_{2}$ conversion factor, $D$ is the distance in megaparsec, and $S_{\mathrm{CO}}$ is the integrated flux of CO in the unit of Jy km s$^{-1}$.
We used the Galactic $X_{\mathrm {CO}}$ of 1.8 $\times$ 10$^{20}$ cm$^{-2}$ (K km s$^{-1}$)$^{-1}$ \citep{Dame01}.
Recent studies have reported that the $X_{\mathrm {CO}}$ factor varies both within a galaxy and between galaxies.
This issue will be discussed in section \ref{Xco}.

The derivation of atomic gas mass is done assuming an optically thin emission, 
\begin{equation}
	M_\mathrm{{H_{I}}}[M_{\odot}] = 2.36 \times 10^{5} D^{2} S_{\mathrm{H\emissiontype{I}}},
\end{equation} 
where $S_{\mathrm{H\emissiontype{I}}}$ is the integrated flux of H\emissiontype{I} in the unit of Jy km s$^{-1}$.

In order to get rid of the effects of the size of progenitors, we normalise the interstellar gas mass with \textit{K$_{s}$}-band luminosity.
Although previous studies such as \citet{Casoli98} and \citet{Zhu01} have used \textit{B}-band luminosity for the normalisation because of data availabilities,
\textit{B}-band severely affected by the emission from young stars and dust.
Therefore, it is not good to use \textit{B}-band for the purpose of the normalisation.
Since \textit{K$_{s}$}-band is less affected by star-formation history than \textit{B}-band \citep{KC98}, we adopted \textit{K$_{s}$}-band luminosity for the normalisation.
The effective wavelength of 2MASS \textit{K$_{s}$}-band, whose data is used for the normalisation, is 2.16 $\mu$m.

For interacting galaxies, we divided the interacting system into two constituent galaxies except for VV 254.
VV 254 is split into three components: UGC 12914, UGC 12915, and the bridge region.
These divisions are performed in the same manner described in Paper I.

The basic properties of interacting galaxies are summarised in table \ref{IG} as following manners:
\begin{quote}
	Column (1): Pair name\\
	Column (2): Galaxy name\\
	Column (3): Molecular hydrogen gas mass assuming the Galactic $I_{\mathrm{CO}}-N_{\mathrm{H2}}$ conversion factor of 1.8 $\times$ 10$^{20}$ [cm$^{-2}$ (K km s$^{-1}$)$^{-1}$]\\
	Column (4): Atomic hydrogen gas mass assuming an optically thin emission, $M_\mathrm{{H_{I}}}[M_{\odot}] = 2.36 \times 10^{5} D^{2} S_{\mathrm{H\emissiontype{I}}}$, where $D$ is the distance in megaparsec, and $S_{\mathrm{H\emissiontype{I}}}$ is the integrated flux of H\emissiontype{I} in Jy km s$^{-1}$\\
	Column (5): H$\alpha$ luminosity. For Arp 84 and VV 254, FUV luminosity is also listed\\
	Column (6): 24 $\mu$m luminosity. For Arp 84, 8 $\mu$m luminosity\\
	Column (7): \textit{K$_{s}$} luminosity\\
	Column (8): Star formation rate\\
	Column (9): Global molecular gas fraction\\
	Column (10): References of H\emissiontype{I}, H$\alpha$, FUV, 24 $\mu$m, and 8 $\mu$m
\end{quote}
Similarly, basic properties of control sample galaxies are summarised in table \ref{COatlas} in the following way:
\begin{quote}
	
	Column (1): Galaxy name\\
	Column (2): Molecular hydrogen gas mass\\
	Column (3): Atomic hydrogen gas mass\\
	Column (4): \textit{K$_{s}$} luminosity\\
	Column (5): Global Molecular gas fraction\\
	Column (6): References of H\emissiontype{I}
\end{quote}
$M_\mathrm{{H_{I}}}$, $L_{\mathrm {H\alpha}}$, $L_{\mathrm{FUV}}$, $L_{\mathrm {24\mu m}}$, $L_{\mathrm {8\mu m}}$, and $L_{K_{s}}$ are calculated based on the distance described in table \ref{isolate}.

\begin{table*}
	\tbl{Basic properties of interacting galaxies}{
		\begin{tabular}{cccccccccc}
			\hline
			Pair name &	Galaxy &  $M_{\mathrm{H_{2}}}$ & $M_{\mathrm{H\emissiontype{I}}}$ &  $L_\mathrm{{H\alpha}}$($L_\mathrm{{FUV}}$) & $L_{24\mu m}$($L_{8\mu m}$)\footnotemark[$*$] & $L_{K_{s}}$ & SFR & $f_{\mathrm{mol}}^{\mathrm{global}}$ & Ref.\footnotemark[$\dagger$] \\
			&  & (10$^{9} \ M_{\odot}$)&   (10$^{9} \ M_{\odot}$) &  (10$^{40(42)}$ erg s$^{-1}$) &   (10$^{42}$ erg s$^{-1}$) & (10$^{43}$ erg s$^{-1}$)  & ($\MO$ yr$^{-1}$) & &\\
			(1)    & (2)           & (3)            & (4)           &  (5)       & (6)    &(7) &(8)& (9) & (10)\\
			\hline 
			Arp 84 & NGC 5394   	& 4.8		&	0.08		& 1.6 (5.1)	& (2.9)		&	3.2		&	4.9  		& 0.98		& 1, 3, 4, 8 \\
			& NGC 5395   	& 8.0		&	5.7		& 1.0 (3.1)	& (10.7)		&	10.8		&	3.1	  	& 0.47		& 1, 3, 4, 8 \\ \hline 
			VV 219 & NGC 4567   	& 1.3		&	0.2		& 3.8			& 0.8  		&	0.67		& 	0.67		& 0.87		& 1, 5, 8 \\
			& NGC 4568   	& 3.0		&	1.1		& 7.9			& 2.0  		& 	2.1		&	1.5		& 0.73		& 1, 5, 8 \\ \hline
			VV 254 &  UGC 12914	& 10.3		&	6.2		& 3.3 (5.1)			& 8.3  		&	8.3		&	1.6		& 0.63		& 1, 4, 6, 9 \\ 
			& UGC 12915	& 14.9		&	5.1		& 2.1 (3.1)			& 4.9		&	4.7		&	4.2		& 0.74		& 1, 4, 6, 9 \\ \hline
			The Antennae Galaxies &NGC 4038  		& 6.2		&	2.5		& 32.9			& 2.7		&	3.3		& 	8.4		& 0.60		& 2, 7, 9 \\ 
			& NGC 4039   	& 3.2		&	1.0		& 30.9			& 2.1		&	2.9		&	9.5		& 0.69		& 2, 7, 9 \\ 
			\hline
			\end{tabular}}
	\label{IG}
	\begin{tabnote}
						\footnotemark[$*$] $L_{24\mu m}$ and $L_{8\mu m}$ are expressed as $\nu L_{\nu}$.\\
						\footnotemark[$\dagger$] Reference: 1.\citet{Iono05}; 2.\citet{Hibbard01}; 3. \citet{Epinat08}; 4. GALEX NGS \citep{Bianchi03};
						5.\citet{Koopmann01}; 6. \citet{Bushouse87}; 7. \citet{Xu00}; 8. \citet{Smith07}; 9. Spitzer archive.
	\end{tabnote}
\end{table*}

\begin{table*}
		\tbl{Basic properties of control isolated spiral galaxies}{
		\begin{tabular}{cccccc}
			\hline
			Galaxy &  $M_{\mathrm{H_{2}}}$ & $M_{\mathrm{H\emissiontype{I}}}$ & $L_{K_{s}}$ & $f_{\mathrm{mol}}^{\mathrm{global}}$ &  Ref.\footnotemark[$*$] \\
			& (10$^{9} \ M_{\odot}$)&   (10$^{9} \ M_{\odot}$) &  (10$^{43}$ erg s$^{-1}$)  &  & \\
			(1)    & (2)           & (3)            & (4) & (5)	&	(6)	\\
			\hline
			NGC 253		 &  1.6		 &   1.0    &     1.9        & 0.49   &   1\\
			UGC 2855	&	20.3   &	8.8		&	5.0			&	0.70 &	2\\
			NGC 3184	&	1.1		&	1.9		&	0.55     	&	0.36 &	3\\
			NGC 3351	&	1.1		&	1.2		&	1.5			&	0.48 &	3\\
			NGC 3521	&	2.9		&	3.6		&	2.2			&	0.44 &	3\\
			NGC 3627	&	8.8		&	1.2		&	5.0			&	0.88 &	3\\
			NGC 3631	&	3.6		&	5.7		&	2.8			&	0.39 &	4\\
			NGC 4051	&	2.7		&	1.9		&	1.6			&	0.59 &	5\\
			NGC 4102	&	2.2		&	0.55	&	2.2			&	0.80 & 6\\
			NGC 5055	&	2.9		&	4.6		&	2.4			&	0.38 &	3\\
			NGC 5248	&	8.8		&	9.9		&	4.8			&	0.47 &	7\\
			NGC 5457	&	3.2		&	13.4	&	1.7			&	0.19 &	3\\
			NGC 6217	&	1.8		&	7.7		&	2.4			&	0.19 &	8\\
			NGC 6951	&	5.9		&	4.8		&	8.3			&	0.55 &	7\\
			\hline
			\end{tabular}}
		\label{COatlas}
		\begin{tabnote}
						\footnotemark[$*$] Reference: 1. \citet{PC91}; 2. \citet{Lang03}; 
						3.\citet{Walter08}; 4. \citet{Knapen97}; 5. \citet{LD95};
						6. \citet{VS01}; 7. \citet{Haan08}; 8. \citet{vanDrielButa91}
		\end{tabnote}
\end{table*}

\begin{figure*}
	\centering
		\FigureFile(160mm,80mm){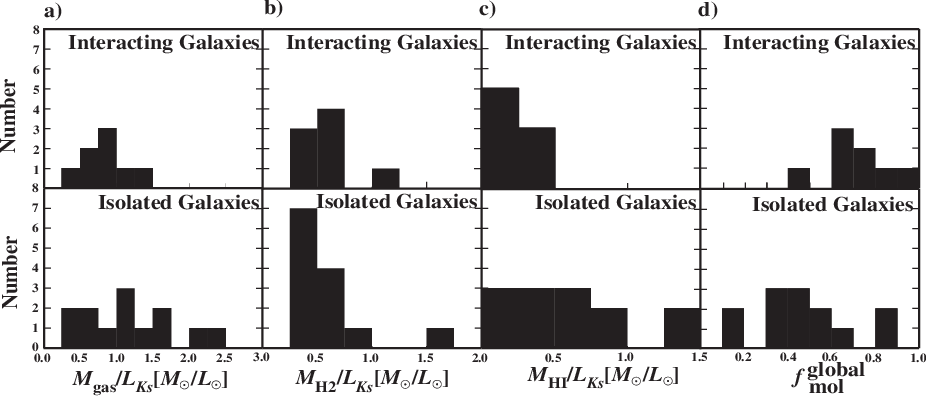}
	\caption{Histograms of  a) $M_{\mathrm{gas}}$/$L_{K_{s}}$, b) $M_{\mathrm{H_{2}}}$/$L_{K_{s}}$, c) $M_{\mathrm{H\emissiontype{I}}}$/$L_{K_{s}}$, and d) $f_{\mathrm{mol}}^{\mathrm{global}}$ for the interacting galaxies (upper figures) and the control sample (lower figures), respectively.}
	\label{histograms}
\end{figure*}

First, we examine a total gas mass (the sum of H$_{2}$ and H\emissiontype{I} gas mass) normalised by {\itshape K$_{s}$}-band luminosity in galactic scale, $M_{\mathrm{gas}}$/$L_{K_{s}}$.
The histogram of  $M_{\mathrm{gas}}$/$L_{K_{s}}$ is shown in figure \ref{histograms}(a).
The average of $M_{\mathrm{gas}}$/$L_{K_{s}}$ for the interacting galaxies and the control sample are 0.90 $\pm$ 0.29 and 1.16 $\pm$ 0.50, respectively.
The medians of $M_{\mathrm{gas}}$/$L_{K_{s}}$ are 0.81 and 1.15 for the interacting galaxies and the control sample, respectively.
Although the average and the median for the interacting galaxies are lower than that for the control sample, the result implies that the normalised total gas mass in the interacting galaxies and the control sample galaxies are the same considering the large dispersion.

Next, we investigate normalised molecular gas mass, $M_{\mathrm{H_{2}}}$/$L_{K_{s}}$.
The histogram of $M_{\mathrm{H_{2}}}$/$L_{K_{s}}$ is shown in figure \ref{histograms}(b).
The average of $M_{\mathrm{H_{2}}}$/$L_{K_{s}}$ for the interacting galaxies and the control sample are 0.67 $\pm$ 0.23 and 0.57 $\pm$ 0.33, respectively.
The medians of $M_{\mathrm{H_{2}}}$/$L_{K_{s}}$ are 0.64 and 0.49 for the interacting galaxies and the control sample, respectively.
The normalised molecular gas mass in the interacting galaxies agrees with that in the isolated galaxies within the error.

We also examine the normalised atomic gas mass.
Figure \ref{histograms}(c) shows the histogram of $M_{\mathrm{H\emissiontype{I}}}$/$L_{K_{s}}$.
For the atomic hydrogen gas mass, interacting galaxies have the average of $M_{\mathrm{H\emissiontype{I}}}$/$L_{K_{s}}$ of 0.23 $\pm$ 0.09 whereas 
$M_{\mathrm{H\emissiontype{I}}}$/$L_{K_{s}}$ of the control sample is 0.59 $\pm$ 0.38.
The median of $M_{\mathrm{H\emissiontype{I}}}$/$L_{K_{s}}$ is 0.24 for interacting galaxies and 0.64 for the control sample.
The dispersion of $M_{\mathrm{H\emissiontype{I}}}$/$L_{K_{s}}$ in isolated galaxies is much larger than that in interacting galaxies.
Furthermore, there is no interacting system whose $M_{\mathrm{H\emissiontype{I}}}$/$L_{K_{s}}$ is higher than 0.50, while seven of 13 (54 \%) isolated spiral galaxies have $M_{\mathrm{H\emissiontype{I}}}$/$L_{Ks}$ of over 0.5, and the highest is 1.34.
These results imply that normalised atomic hydrogen gas mass could be decreased during interactions.

\section{Discussion}
\subsection{Global molecular gas fraction}
\label{globalfmol}
In interstellar space, a transition from atomic hydrogen gas to molecular hydrogen gas can occur on the surface of dust. 
In order to quantify the balance of the transition, a global molecular gas fraction to total gas, $f_{\mathrm{mol}}^{\mathrm{global}}$, is used (e.g., \cite{Elmegreen93}).
The definition of $f_{\mathrm{mol}}^{\mathrm{global}}$ is given as follows:
\begin{equation}
	f_{\mathrm{mol}}^{\mathrm{global}}=\frac{M_{\mathrm{H_{2}}}}{M_{\mathrm{H\emissiontype{I}}}+{M_{\mathrm{H_{2}}}}}.
\end{equation}

If the transition from atomic gas to molecular gas is more efficient in interacting galaxies than in isolated spiral galaxies,
a comparison of $f_{\mathrm{mol}}^{\mathrm{global}}$ should give us suggestive information to see the influence of the interaction on the ISM.
Figure \ref{histograms}(d) shows histograms of $f_{\mathrm{mol}}^{\mathrm{global}}$ for both the interacting galaxies and the control galaxies samples.
Interacting galaxies have a higher value of $f_{\mathrm{mol}}^{\mathrm{global}}$ compared to the isolated spiral galaxies sample.
The averages of interacting galaxies and the control sample are 0.71 $\pm$ 0.15 and 0.52 $\pm$ 0.18, respectively.
The medians of $f_{\mathrm{mol}}^{\mathrm{global}}$ are 0.71 and 0.48 for interacting galaxies and the control sample, respectively. 
This fact suggests that molecular hydrogen gas is produced from atomic hydrogen gas effectively in interacting galaxies.
The global molecular gas fraction differs according to the Hubble type, that is, early-type galaxies tend to have higher global molecular gas fraction compared to late-type galaxies \citep{YK89}.
However, since progenitors of our target interacting galaxies are all late-type galaxies based on the RC3 Catalogue \citep{RC3}, the high global molecular gas fraction cannot be explained by the difference of the Hubble type.
Therefore, these results imply that the galaxy interactions do enhance the global molecular gas fraction in a galaxy even in the early stage of the interaction.

It is known that H\emissiontype{I} gas may be selectively stripped under group or cluster environments \citep{KY89}.
We have checked whether H\emissiontype{I} gas apart from the galaxies exists in the H\emissiontype{I} integrated intensity map (see, figure~6(c)--9(c) in Paper I) in order to investigate a possibility of higher $f_{\mathrm{mol}}^{\mathrm{global}}$ due to H\emissiontype{I} gas removal.
We find no atomic gas far from the disk of the galaxies.
Thus, we did not underestimate the atomic gas mass.
We also consider the current environments.
Since the VV 219 system (NGC 4567 and NGC 4568) is in the Virgo Cluster, ram pressure may selectively strip H\emissiontype{I} gas only in this system in our sample.
Even if we remove NGC 4567 and NGC 4568 from the sample of the interacting galaxies, the average (0.68 $\pm$ 0.17) and the median (0.65) hold our conclusion of higher $f_{\mathrm{mol}}^{\mathrm{global}}$ in the interacting galaxies.
Therefore, the enhancement of $f_{\mathrm{mol}}^{\mathrm{global}}$ is not from environmental effects but due to the nature of a galaxy interaction itself. 

Although we cannot find stripped H\emissiontype{I} gas apart from progenitors, there is still a possibility of ionisation of such stripped H\emissiontype{I} gas.
If it exists, it will be another candidate to enhance $f_{\mathrm{mol}}^{\mathrm{global}}$.	
Once the H\emissiontype{I} gas is ionised by a local H\emissiontype{II} region, the H\emissiontype{I} gas disappears, and dense  ionised gas can be seen at the ionised region \citep{Hibbard96}.
In this context, we examined the distribution of ionised gas with H$\alpha$ image and of ionising sources with FUV image where molecular gas is not seen as the case of the tidal tail of the Antennae Galaxies \citep{Hibbard01}.
NGC 5394 (the north-east galaxy in Arp 84) shows extended FUV emission along their arms without H\emissiontype{I} emission.
Although this feature may be due to an interaction-induced star formation, we can say that ionisation of hydrogen gas is at least occurring there.
In other galaxies, we can find no H$\alpha$ or FUV emission apart from their disc even in VV 254, from which a significant amount of H\emissiontype{I} gas has dragged out (see, Paper I: figure~6(e) -- 9(e)).
This fact indicates that the current ionisation of atomic gas can be negligible except for NGC 5394, although past ionisation due to the interaction cannot be ruled out.

We also derive H\emissiontype{II} gas mass outside galactic discs in order to evaluate a contribution of warm diffuse ionised gas to H\emissiontype{I} deficiency in interacting galaxies.
Diffuse H$\alpha$ emission traces warm ionised gas having a temperature of $\sim$10$^{4}$ K and electron density of lower than 10$^{3}$ cm$^{-3}$.
In this analysis, a size of a galactic disc is defined as a radius within 20 mag arcsec$^{-2}$ in the {\itshape Ks}-band ($R_{K20}$).
We regard H$\alpha$ emission within 20 kpc as whole emission from ionised gas in a galaxy.
To obtain diffuse H$\alpha$ emission outside galactic discs, we subtract H$\alpha$ flux within $R_{K20}$ from that within 20 kpc.
Assuming an electron temperature of $\sim$ 10$^{4}$ K, H\emissiontype{II} gas mass can be derived using the following equation \citep{Finkelman10}:
\begin{equation}
	M_{\mathrm{HII}} [M_{\odot}]= 2.33\times10^{3}\left(\frac{L_{\mathrm{H}\alpha}}{10^{39} \ \mathrm{erg \ s^{-1}}}\right)  \left(\frac{n_{e}}{10^{3} \ \mathrm{cm^{-3}}}\right)^{-1},
\end{equation}
where $n_{e}$ is the electron density. 
\citet{Krabbe14} showed that interacting galaxies in the early stage have higher electron density than isolated galaxies.
The electron density stays $>$ a few 10 cm$^{-3}$ even further than 10 kpc from the galactic centre in some interacting galaxies.
We adopt the electron density of 10 cm$^{-3}$ to estimate H\emissiontype{II} gas mass.
Table \ref{ionisedgas} shows derived H\emissiontype{II} gas mass.
All target interacting galaxies expect for NGC 5394 have an order of 3 smaller H\emissiontype{II} gas mass compared with H\emissiontype{I} gas mass.
It means that an ionisation of hydrogen gas can be negligible for these interacting galaxies except for NGC 5394 as suggested by distributions of H$\alpha$ and FUV emission.

H\emissiontype{I} gas expelled from galactic discs may be suffered by hot halo or inter-galactic UV radiation which cannot be detected with H$\alpha$ emission.
Although the origin of hot diffuse ionised gas is unknown yet \citep{Werk16}, H\emissiontype{I} may become hot diffuse ionised gas with a temperature of $\sim$10$^{5-6}$ K and electron density of $<$1 cm$^{-3}$, and consist circum-galactic medium which extends over several 100 kpc as reported in \citet{Werk14}.
In 25\% of massive local early-type galaxies, a misalignment is seen between a position angle derived from warm ionised gas and that from with stellar kinematics \citep{Pandya17}.
Such a misalignment of the hot ionised gas and stellar discs indicates that the disc of the hot ionised gas is formed by an external process.
As shown by \citet{vandeVoort15}, merger-induced misalignment lasts 2 Gyr.
These facts suggest that at least some early-type galaxies acquire their diffuse warm ionised gas via a merger event.
Additionally, according to table 3 of \citet{HM12}, a photoionisation rate owing to the cosmic background UV emission at z = 0 is 0.228$\times$10$^{-13}$ s$^{-1}$ (i.e., a photoionisation timescale is an order of 10$^{6}$ yr). 
Since this timescale is shorter than the timescale of the end of the first encounter ($\sim$10$^{8}$ yr: e.g., \cite{TCB10}), H\emissiontype{I} gas in a tidal tail can be photoionised.
Although there is no data about hot diffuse ionised gas for our sample, it is possible that some portion of H\emissiontype{I} gas in the tidal tail in our sample could be transformed into hot diffuse ionised gas. 
Therefore, we conclude that higher global molecular fraction in interacting galaxies is achieved by both production of molecular gas from atomic gas and past ionisation of atomic gas.

\begin{table*}[tbp]
	\centering
		\tbl{H\emissiontype{II} gas mass around galactic discs of interacting galaxies}{
		\begin{tabular}{ccccc}
			\hline
			Pair name&Galaxy & H$\alpha$ fux within {\itshape R}$_{K20}$ & H$\alpha$ flux within 20 kpc & H\emissiontype{II} gas mass  outside the galactic disc \\
			& & (10$^{40}$ erg s$^{-1}$) & (10$^{40}$ erg s$^{-1}$)  & (10$^{6}$ {\itshape M}$_{\odot}$) \\ 
			\hline
			Arp 84&NGC 5394 & 0.4 & 1.6  & 2.8 \\
			&NGC 5395 & 0.4 & 1.0 & 1.4 \\
			\hline
			VV 219&NGC 4567 & 3.9 & 4.3  & 0.9 \\
			&NGC 4568 & 8.9 & 10.4  & 3.4 \\
			\hline
			VV 254&UGC 12914 & 2.8 & 3.3 & 1.2\\
			&UGC 12915 & 1.6 & 2.1  & 1.2\\
			\hline
			The Antennae Galaxies & NGC 4038 & 31.0 & 32.9 & 4.3 \\
												 & NGC 4039 & 27.7 & 30.9 & 7.4\\
			\hline
		\end{tabular}}
		\label{ionisedgas}
\end{table*}

\subsection{Local molecular gas fraction}
\label{localfmol}
In order to make it clear where and how molecular gas fraction is enhanced in interacting galaxies, 
we investigate spatial variations of a local molecular gas fraction.
Local molecular gas fraction is derived using the following equation:
\begin{equation}
	f_{\mathrm{mol}}=\frac{\Sigma_{\mathrm{H_{2}}}}{\Sigma_{\mathrm{H\emissiontype{I}}}+\Sigma_{\mathrm{H_{2}}}},
\end{equation}
where $\Sigma_{\mathrm{ H_{2}}}$ is the surface density of molecular hydrogen gas and $\Sigma_{\mathrm{H\emissiontype{I}}}$ is the surface density of atomic hydrogen gas.
$\Sigma_{\mathrm{ H_{2}}}$ and $\Sigma_{\mathrm{H\emissiontype{I}}}$ are derived using the following equations:

\begin{equation}
	\frac{\Sigma_{\mathrm{H_{2}}}}{M_{\odot} \ \mathrm{pc}^{-2}} =
	2.88  \left(\frac{X_{\mathrm{CO}}}{X_{\mathrm{CO}}^{\mathrm{Galactic}}}\right) \left(\frac{I_{\mathrm{CO}}}{\mathrm{K \ km \ s^{-1}}}\right) \mathrm{cos} \ i_{\mathrm{CO}},	\label{h2mass}
\end{equation}

\begin{equation}
	\frac{\Sigma_{\mathrm{H\emissiontype{I}}}}{M_{\odot} \ \mathrm{pc}^{-2}} = 
	23.5  \left(\frac{I_{\mathrm{H\emissiontype{I}}}}{\mathrm{Jy \ km \ s^{-1}}}\right) \mathrm{cos} \  i_{\mathrm{H\emissiontype{I}}} 	\label{h1mass},
\end{equation}
where $i_{\mathrm{CO}}$ and $i_{\mathrm{H\emissiontype{I}}}$ are the inclination angle of the galaxy derived from CO and H\emissiontype{I} data.
$X_{\mathrm{CO}}^{\mathrm{Galactic}}$ is the Galactic CO-H$_{2}$ conversion factor of $1.8\times 10^{20} \mathrm{cm^{-2} \ (K \ km \ s^{-1})^{-1}}$ \citep{Dame01}.
In this paper, these inclinations are assumed to be the same throughout the disc for each galaxy and between CO disc and H\emissiontype{I} disc.
Note that as long as the inclinations of H\emissiontype{I} disc and H$_{2}$ disc are the same, 
and $f_{\mathrm{mol}}$ is free from the problem of the uncertainty of the inclination of galaxies.
If the H$_{2}$ disc is inclined more than what we assumed and the inclination angle of H\emissiontype{I} disc is the same as our assumption, calculated $f_{\mathrm{mol}}$ gets lower than the real and vice versa. 
We also notice that we derive $f_{\mathrm{mol}}$ for only pixels where CO emission is detected higher than 3 $\sigma$.
For this reason, this investigation can check local molecular-to-atomic gas balance.

\begin{figure}[tbp]
	\centering
		\FigureFile(80mm,50mm){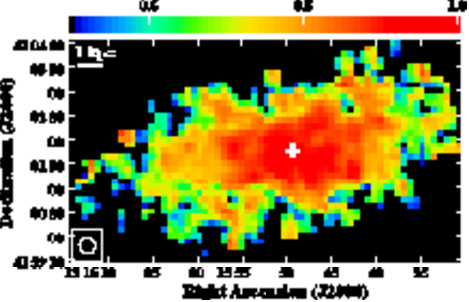}
		\caption{A map of $f_{\mathrm {mol}}$ of a typical isolated spiral galaxy NGC 5055. The white cross shows the centre of the galaxy.}
		\label{NGC5055fmol}
\end{figure}
\begin{figure*}[htbp]
	\centering
		\FigureFile(158mm,165mm){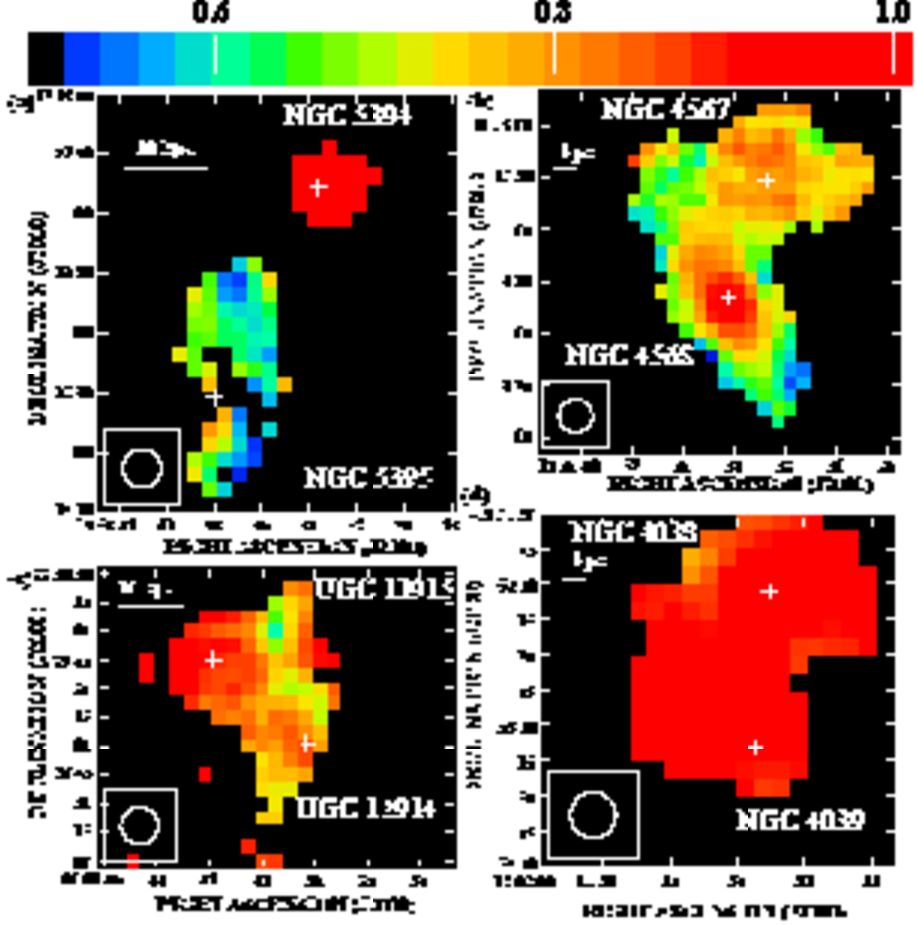}
		\caption{Maps of $f_{\mathrm {mol}}$ for the interacting galaxies. (a) Arp 84, (b) VV 219, (c) VV 254, and (d) the Antennae Galaxies. The crosses illustrate the galactic centre of each constituent galaxy. White circles show the beam size of NRO 45-m (\timeform{19''.3}) on bottom left of each figure. White line on each top left of the figures represents linear scale.}
		\label{fmol}
\end{figure*}

First of all, we compare the distributions of $f_{\mathrm{mol}}$ between isolated galaxies and interacting galaxies.
Figure \ref{NGC5055fmol} is a map of $f_{\mathrm{mol}}$ of NGC 5055, which is in the control sample.
This map illustrates an example of the typical $f_{\mathrm{mol}}$ distribution seen in isolated galaxies.
$f_{\mathrm{mol}}$ has a peak at the galactic centre where the peak surface density of molecular gas is also located (see, figure 33 in \cite{Kuno07}) and gradually decreases as the radius increases.
Maps of $f_{\mathrm{mol}}$ in the interacting galaxies are shown in figure \ref{fmol}.
Note that all maps of $f_{\mathrm{mol}}$ in figures \ref{NGC5055fmol} and \ref{fmol} are set to have the same dynamic range.
These figures illustrate that the distributions of $f_{\mathrm{mol}}$ in interacting galaxies are complicated. 
Here, we see the features of $f_{\mathrm{mol}}$ map of each galaxy pair.

A map of $f_{\mathrm{mol}}$ of Arp 84 (figure \ref{fmol}(a)) displays that hydrogen gas in NGC 5394 (top-right) is almost entirely in a molecular state throughout the galaxy.
NGC 5395 (bottom-left) has a local peak of $f_{\mathrm{mol}}$ at the centre where CO and H\emissiontype{I} maps show cavity-like structures (see, figure 6(c) in Paper I).
The edge of the northern tidal arm of NGC 5395 toward NGC 5394 where molecular gas and atomic gas are accumulated as mentioned in Paper I shows slightly higher $f_{\mathrm{mol}}$ which is comparable to the galactic centre.
This local peak of $f_{\mathrm{mol}}$ comes into contact with the optical tidal tail of NGC 5394.
These complicated distributions of $f_{\mathrm{mol}}$ indicate that the galaxy interactions have had a large influence on the properties of the interstellar gas.

As seen in figure \ref{fmol}(b), both NGC 4567 (top-right) and NGC 4568 (bottom-left) which are the constituent galaxies of VV 219 show a relatively ordinary distribution of $f_{\mathrm{mol}}$ as found in normal field galaxies such as NGC 5055 (figure \ref{NGC5055fmol}).
They have peaks at their centres and a gradual decrease in $f_{\mathrm{mol}}$ with galactocentric radius.
However, $f_{\mathrm{mol}}$ in NGC 4567 decreases along with an increase of the radius more gradually than NGC 4568 and NGC 5055.
There is a slight enhancement of $f_{\mathrm{mol}}$ at the overlapping region of the galaxies.

VV 254 shows the most complicated distribution of $f_{\mathrm{mol}}$ among our sample (figure \ref{fmol}(c)).
There are two peaks of $f_{\mathrm{mol}}$ in the VV 254 system.
One is about \timeform{45''} (corresponding 14 kpc in projected distance) north from the centre of UGC 12914 (bottom-right), which is the edge of the warped disc of UGC 12914.
Another is about \timeform{20''} (6 kpc) to the east from the centre of UGC 12915.
From this highest $f_{\mathrm{mol}}$ region in VV 254 to the centre of UGC 12914, $f_{\mathrm{mol}}$ gradually decreases.
The lowest $f_{\mathrm{mol}}$ ($\sim$ 0.6) of this system is seen in the northwestern region of UGC 12915.
Except for this lowest region, VV 254 system has $f_{\mathrm{mol}}$ over 0.7 throughout the galaxy even at the bridge region.

Figure \ref{fmol}(d) illustrates that $f_{\mathrm{mol}}$ in the Antennae Galaxies is almost over 0.9 throughout the region where CO emission is detected.
\begin{figure}[tbp]
	\centering
		\FigureFile(80mm,55mm){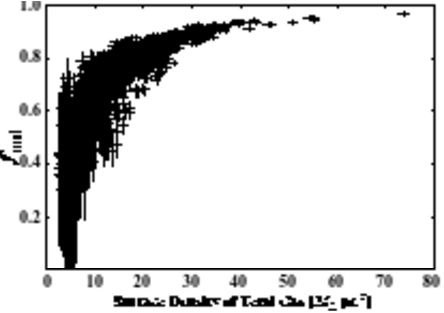}
		\caption{$f_{\mathrm {mol}}$ versus surface density of total gas ($\Sigma_{\mathrm{gas}}$) for isolated galaxy NGC 5055.}
		\label{NGC5055fmolmass}
\end{figure}

In order to see the relation more quantitatively, we made plots of the relation between the surface density of the total gas and $f_{\mathrm{mol}}$. 
The surface density of the total gas is written as $\Sigma_{\mathrm{gas}} = \Sigma_{\mathrm{H_{2}}}+\Sigma_{\mathrm{H\emissiontype{I}}}$. 
We made the correction assuming that the inclination angle is constant within a galaxy.
It should be noted that $f_{\mathrm{mol}}$ is not affected by the inclination of the galaxy as previously mentioned,
while it leads to uncertainty of the surface density of total gas if the inclination that we assume is not true.
For both CO and H\emissiontype{I} detected points, we estimate errors of $f_{\mathrm{mol}}$ and $\Sigma_{\mathrm{gas}}$ using the errors on the intensity of CO and H\emissiontype{I}.
The errors on the intensity of CO and H\emissiontype{I} are evaluated as follows (e.g., \cite{Sauty03}):
\begin{eqnarray}
\Delta I_{\mathrm{CO}}=\sigma_{\mathrm{CO}}(\Delta V_{\mathrm{H\emissiontype{I}}}\delta V_{\mathrm{CO}})^{1/2}\\
\Delta I_{\mathrm{H\emissiontype{I}}}=\sigma_{\mathrm{H\emissiontype{I}}}(\Delta V_{\mathrm{CO}}\delta V_{\mathrm{H\emissiontype{I}}})^{1/2},
\end{eqnarray}
where $\sigma_{\mathrm{CO}}$ and $\sigma_{\mathrm{H\emissiontype{I}}}$ are the rms noise of the spectrum, 
$\Delta V_{\mathrm{CO}}$ and $\Delta V_{\mathrm{H\emissiontype{I}}}$ linewidth of CO and H\emissiontype{I}, 
$\delta V_{\mathrm{CO}}$ and $\delta V_{\mathrm{H\emissiontype{I}}}$ the velocity resolution of CO and H\emissiontype{I}.
The error of $f_{\mathrm{mol}}$ was expressed as
\begin{equation}
\Delta f_{\mathrm{mol}}= \sqrt{\left(\frac{\partial f_{\mathrm{mol}}}{\partial \Sigma_{\mathrm{H_{2}}}}\Delta\Sigma_{\mathrm{H_{2}}}\right)^{2}
	+\left(\frac{\partial f_{\mathrm{mol}}}{\partial \Sigma_{\mathrm{H\emissiontype{I}}}}\Delta\Sigma_{\mathrm{H\emissiontype{I}}}\right)^{2}},
\end{equation}
where
\begin{eqnarray}
\frac{\partial f_{\mathrm{mol}}}{\partial \Sigma_{\mathrm{H_{2}}}} &=&\frac{\Sigma_{\mathrm{H\emissiontype{I}}}}{(\Sigma_{\mathrm{H\emissiontype{I}}}+\Sigma_{\mathrm{H_{2}}})^{2}}\\
\frac{\partial f_{\mathrm{mol}}}{\partial \Sigma_{\mathrm{H\emissiontype{I}}}} &=& -\frac{\Sigma_{\mathrm{H_{2}}}}{(\Sigma_{\mathrm{H_{1}}}+\Sigma_{\mathrm{H_{2}}})^{2}}.
\end{eqnarray}
$\Delta\Sigma_{\mathrm{H_{2}}}$ and $\Delta\Sigma_{\mathrm{H\emissiontype{I}}}$ are derived from the equation (\ref{h2mass}) and (\ref{h1mass}) 
applying $\Delta I_{\mathrm{CO}}$ and $\Delta I_{\mathrm{H\emissiontype{I}}}$ instead of $I_{\mathrm{CO}}$ and $I_{\mathrm{H\emissiontype{I}}}$, respectively. 

We show $\Sigma_{\mathrm{gas}}$-$f_{\mathrm{mol}}$ plots with error bars.
For comparison, we also made a $\Sigma_{\mathrm{gas}}$-$f_{\mathrm{mol}}$ plot for the isolated spiral galaxy NGC 5055 (figure \ref{NGC5055fmolmass}).
The plot for NGC 5055 shows the gradual increase of $f_{\mathrm{mol}}$ with the surface density of the total gas toward a saturation level of 1.0.
This is a typical feature seen in normal galaxies \citep{Nakanishi06, Tosaki11}.

\begin{figure*}[tbp]
	\centering
		\FigureFile(160mm,115mm){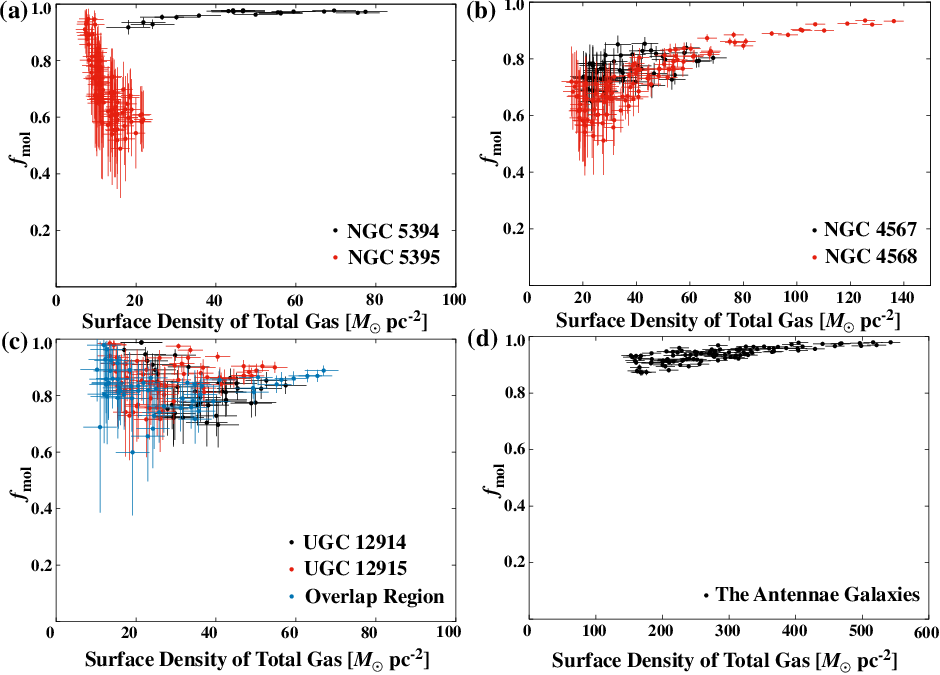}
		\caption{$f_{\mathrm {mol}}$ versus surface density of total gas ($\Sigma_{\mathrm{gas}}$) for (a) Arp 84, (b) VV 219, (c) VV 254, and (d) the Antennae Galaxies.}
		\label{fmolmassErr}
\end{figure*}

Figure \ref{fmolmassErr}(a) shows the plot of the surface density of total gas versus $f_{\mathrm{mol}}$ in Arp 84.
As shown in figure \ref{fmol}(a), NGC 5394 has $f_{\mathrm{mol}}$ of nearly 1.0 throughout the galaxy even in the region with a low surface density of the total gas.
On the other hand, NGC 5395 shows a trend opposite that of  normal galaxies, namely, $f_{\mathrm{mol}}$ decreases with the surface density of the total gas.
The same trend in other galaxies has never yet been reported.

Figure \ref{fmolmassErr}(b) represents the plot of the surface density of the total gas versus $f_{\mathrm{mol}}$ in VV 219.
The plot shows that $f_{\mathrm{mol}}$ in NGC 4567 ranges from 0.6 to 0.9 and fairly constant, 
while $f_{\mathrm{mol}}$ increases with the surface density of total gas in NGC 4568 as seen in normal galaxies.
This difference may be due to the difference in the significance of the influence of interaction on these galaxies.
Since NGC 4567 is a secondary galaxy in the minor merger, it may be more severely affected by the interaction than NGC 4568.

Figure \ref{fmolmassErr}(c) shows the plot of the surface density of the total gas versus $f_{\mathrm{mol}}$ in VV 254.
In this galaxy pair, all $f_{\mathrm{mol}}$ is higher than 0.6 in the region where CO emission is detected.
Both UGC 12914 and UGC 12915 show a large dispersion of $f_{\mathrm{mol}}$.
The overlapping region shows slightly lower $f_{\mathrm{mol}}$ than the galaxies. 
None of the regions shows a clear trend.

Figure \ref{fmolmassErr}(d) shows the plot of the surface density of the total gas versus $f_{\mathrm{mol}}$ in the Antennae Galaxies. 
We make no correction for the inclination for the Antennae Galaxies because of their severely disturbed morphologies. 
We also note that data points move to the left in the plot overall if we could make the inclination correction.
Most $f_{\mathrm{mol}}$ of the Antennae Galaxies is at least 0.9 or higher along a wide range of the surface density of the total gas.

All of our samples show different features compared to isolated galaxies. 
One galaxy pair, VV 254, show relatively constant $f_{\mathrm{mol}}$ and large dispersion and $f_{\mathrm{mol}}$ of Arp 84 and the Antennae Galaxies reaches $>$ 0.8 even at a low surface density of the total gas around 20 $\MO$ pc$^{-2}$.
In the extreme case of NGC 5395, $f_{\mathrm{mol}}$ decreases with the surface density of the total gas. 
As a whole, $f_{\mathrm{mol}}$ in interacting galaxies is not merely related to the surface density of the total gas as often seen in isolated spiral galaxies.

\subsection{The theoretical model of $f_{\mathrm{mol}}$ and fitting to the observed $f_{\mathrm{mol}}$}
\label{fmoldiscuss}
We investigate a cause of the features seen in the plot of the surface density of the total gas versus $f_{\mathrm{mol}}$ of the interacting galaxies, by fitting the data with the theoretical model proposed by \citet{Elmegreen93}.
In his model, $f_{\mathrm{mol}}$ is determined by the balance between the production and the destruction of molecular gas.
This means that the model assumes that the H\emissiontype{I} and H$_{2}$ clouds co-exist in the same place all the time. 
It is not clear that the H\emissiontype{I} and H$_{2}$ clouds co-exist in interacting galaxies, especially in a colliding system, although this assumption is valid for a normal spiral galaxy. 
This issue will be discussed in section \ref{mechanismfmol}.

The production rate from H\emissiontype{I} to H$_{2}$ and H$_{2}$ destruction rate are determined by following phenomena.
Molecular hydrogen gas is produced from two atomic hydrogens on a surface of interstellar dust which is a compound of the metal.
The production rate is proportional to the square of the density of hydrogen gas and pressure that governs whether clouds are self-gravitating or diffuse.
Therefore, the production rate is enhanced under rich metallicity and high pressure.
The energy for photodissociation of molecular hydrogen is supplied by UV radiation emitted from nearby OB stars.
The destruction rate of molecular gas decreases when the interstellar dust is abundant because it shields molecular gas inside a molecular cloud from the UV radiation.
Foreground molecular gas clouds toward an OB star also can shield background molecular gas clouds from photodissociation by the UV radiation.
Furthermore, if the density of a molecular cloud is dense enough, an outer layer of the molecular cloud can shield an inner layer of the molecular cloud from the UV radiation.
Thus the photodissociation rate relates to the amount of dust and the density of the molecular cloud.

Although the relation between the pressure and the surface density of gas is still under discussion,
\citet{Elmegreen89} proposed the relation between kinematic pressure in interstellar medium and the surface density of gas as follows: 
\begin{equation}
	P\simeq \frac{\pi}{2}G\Sigma_{\mathrm{gas}}\left(\Sigma_{\mathrm{gas}}+\Sigma_{\mathrm{star}}\frac{c_{\mathrm{gas}}}{c_{\mathrm{star}}}\right),
\end{equation}
where $G$ is the gravitational constant and $\Sigma_{\mathrm{star}}$ the surface density of stars, 
and $c_{\mathrm{gas}}$ and $c_{\mathrm{star}}$ the gas and star velocity dispersions, respectively.
According to the observational fact that both $\Sigma_{\mathrm{gas}}$ and $\Sigma_{\mathrm{star}}$ have about the same exponential decline with radius \citep{YS82},
this relation is simplified with normalisation in the solar neighbourhood as 
\begin{equation}
	\frac{P}{P_{\odot}} =  \left(\frac{\Sigma_{\mathrm{gas}}}{\Sigma_{\mathrm{gas \odot}}}\right)^{2}.
	\label{press}
\end{equation}
Summarising these production and destruction rate, primary parameters for a balance of $f_{\mathrm{mol}}$ are metallicity, $Z$, the strength of UV radiation field, $U$, and the surface density of the total gas, $\Sigma_{\mathrm{gas}}$.  

The strength of radiation field $U$ can be traced by SFR which is calculated using observables, for instance, H$\alpha$ (or FUV) and 24 $\mu$m (8 $\mu$m) data. These methods to derive dust absorption corrected SFR are proposed by \citet{Calzetti07, Zhu08}.
Therefore, we can compute $f_{\mathrm{mol}}$ using a local surface density of total gas $\Sigma_{\mathrm{gas}}$ and local SFR $\Sigma_{\mathrm{SFR}}$ for the pixel by pixel.
For this reason, free parameters are reduced to only one parameter, metallicity $Z$.
In order to normalise parameters, we assume $\Sigma_{\mathrm{gas}}$ at the solar neighbourhood as 8.0 $\MO$ pc$^{-2}$ \citep{SSS84}, and 
$\Sigma_{\mathrm{SFR}}$ at the solar neighbourhood is estimated as 2.70 $\times$ 10$^{-9} \ M_{\odot}$ yr$^{-1}$ pc$^{-2}$ 
which is obtained by \citet{Bigiel08} using the empirical relation between the surface density of SFR and the total gas (Kennicutt-Schmidt law). 

\begin{figure}[tbp]
	\centering
		\FigureFile(80mm,57mm){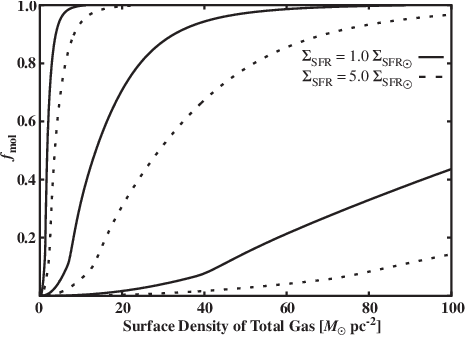}
		\caption{The effects of metallicity and $\Sigma_{\mathrm{SFR}}$ on $f_{\mathrm {mol}}$ versus the surface density of total gas relation.
			Solid and dotted lines indicates fixed $\Sigma_{\mathrm{SFR}}$ = 1.0 $\Sigma_{\mathrm{SFR_{\odot}}}$, and  fixed $\Sigma_{\mathrm{SFR}}$ = 5.0 $\Sigma_{\mathrm{SFR_{\odot}}}$, respectively.
			Focusing on same $\Sigma_{\mathrm{SFR}}$,  three $f_{\mathrm {mol}}$-$\Sigma_{\mathrm{gas}}$ relations from bottom one to top one correspond in the case of metallicity $Z$ of 0.1, 1.0, and 10.0.}
		\label{fmolmetallicity}
\end{figure}

Figure \ref{fmolmetallicity} shows the relationship between simulated $f_{\mathrm {mol}}$ versus the surface density of the total gas with different metallicity $Z$ in case of $\Sigma_{\mathrm{SFR}}$ = 1.0 $\Sigma_{\mathrm{SFR_{\odot}}}$ (solid lines) and $\Sigma_{\mathrm{SFR}}$ = 5.0 $\Sigma_{\mathrm{SFR_{\odot}}}$ (dotted lines).
For each $\Sigma_{\mathrm{SFR}}$, we calculate three cases that metallicity of 0.1, 1.0, and 10.0 which correspond from the bottom line to the top one. 
It is apparent that $f_{\mathrm {mol}}$ decreases when $\Sigma_{\mathrm{SFR}}$ increases.
On the other hand, an increase of $Z$ leads to increase $f_{\mathrm {mol}}$.

It is necessary to consider that normally a galaxy has a radial gradient on metallicity.
$f_{\mathrm{mol}}$ fitting with constant metallicity for a normal spiral galaxy could lead to misinterpretation.
\citet{Kewley10}, however, showed that the metallicity gradients in all interacting galaxies in early stages are significantly shallower than that in the isolated spiral galaxies
(VV 254, which is our sample interacting galaxies, is included in their targets).
Hence, for the interacting galaxies, simulations with constant metallicity throughout a galaxy are less affected than those carried out for isolated galaxies.
Therefore, we fit the observed data with this model assuming that metallicity $Z$ is constant throughout a galaxy for simplification.
In order to constrain parameter sets, we derive metallicity from the luminosity--metallicity relation.
\citet{MS02} showed that $B$-band luminosity is correlated with metallicity as:
\begin{equation}
12 + \mathrm{log (O/H)} = -(0.240 \pm 0.006) M_{B} + (4.059 \pm 0.17),  
\end{equation}
where $M_{B}$ is absolute $B$-band magnitude of a galaxy.
We assume log~$Z_{\odot}$~=~12~+~log (O/H)$_{\odot}$~=~8.69.
We collect total apparent $B$-band magnitude from SIMBAD.
Since the bridge region of VV 254 is not available, the metallicity of the bridge region is not calculated.
Besides, metallicity of the Antennae Galaxies is derived summing up their progenitors, NGC 4038 and NGC 4039.
The derived metallicity $Z/Z_{\odot}$ is summarised in table \ref{metallicity}.
As a result, we get more constrained parameter set for all galaxies except for the bridge region of VV 254.
\begin{table*}[tbp]
	\centering
		\tbl{\itshape{B}-band magnitude and derived metallicity}{
		\begin{tabular}{cccccc}
			\hline
			Pair name&Galaxy & Total $B$-band magnitude & Absolute $B$-band magnitude & log $Z$ & $Z/Z_{\odot}$ \\
			& & (mag) & (mag) &  & \\
			\hline
			Arp 84&NGC 5394 & 13.70 & -19.92 & 8.84 & 1.4$^{2.5}_{0.8}$ \\
			&NGC 5395 & 13.26 & -20.36 & 8.95 & 1.8$^{3.2}_{1.0}$\\
			\hline
			VV 219&NGC 4567 & 12.06 & -18.96 & 8.61 & 0.8$^{1.5}_{0.5}$ \\
			&NGC 4568 & 12.11 & -19.92 & 8.60& 0.8$^{1.5}_{0.5}$ \\
			\hline
			VV 254&UGC 12914 & 13.2 & -20.80 & 9.04 & 2.2$^{4.0}_{1.3}$ \\
			&UGC 12915 & 13.9 & -20.10 & 8.87 & 1.5$^{2.7}_{0.9}$ \\
			\hline
			The Antennae Galaxies && 10.24 & -21.47 & 9.21 & 3.3$^{5.9}_{1.9}$\\
			\hline
		\end{tabular}}
		\label{metallicity}
	
\end{table*}

\begin{figure}[tbp]
	\centering
		\FigureFile(80mm,57mm){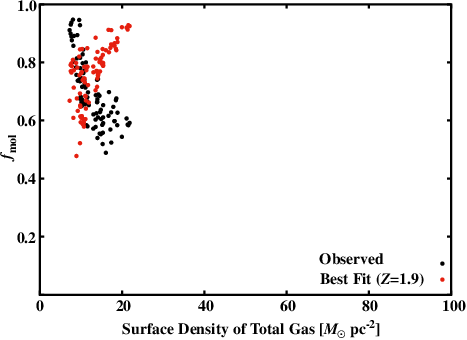}
		\caption{The results of theoretical $f_{\mathrm{mol}}$ model fitting for NGC 5395. 
						In this fitting, free parameter is only metallicity $Z$. 
						Black points are observed $f_{\mathrm{mol}}$, while red points are best-fitted theoretical $f_{\mathrm{mol}}$.}
		\label{ngc5395woExt}
	
\end{figure}

For the first step, we fit NGC 5395, which shows the oddest $f_{\mathrm{mol}}$--$\Sigma_{\mathrm {gas}}$ relation in our sample, to test the simulation validity.
The best-fitted parameter of metallicity is 1.0 $Z_{\odot}$ with minimum $\chi^{2}$ of 4.1.
Figure \ref{ngc5395woExt} shows observed and simulated $f_{\mathrm{mol}}$--$\Sigma_{\mathrm {gas}}$ relations of NGC 5395.
Black and red points correspond observed $f_{\mathrm{mol}}$--$\Sigma_{\mathrm {gas}}$ relation and simulated $f_{\mathrm{mol}}$--$\Sigma_{\mathrm {gas}}$ relation with metallicity of 1.0 $Z_{\odot}$ yielding the minimum $\chi^{2}$ within a range of $Z$, respectively.
It is clear that we could not reproduce the observed trend that $f_{\mathrm{mol}}$ decreases with the increase of the surface density of the total gas. 
Therefore, it needs to modify Elmegreen's original theoretical $f_{\mathrm{mol}}$ model to represent the physics in interacting galaxies.

\begin{figure}[tbp]
	\centering
		\FigureFile(80mm,58mm){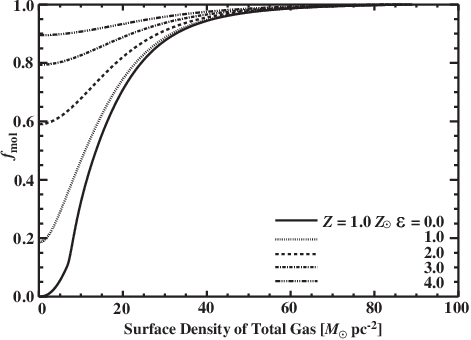}
		\caption{$f_{\mathrm {mol}}$ versus surface density of total gas ($\Sigma_{\mathrm{gas}}$) if external pressure $\varepsilon$ induced.}
		\label{fmolexpressure}
	
\end{figure}

Besides metallicity $Z$, we tried to include external pressure $\varepsilon$ which expresses the influence of the interaction of galaxies as another free parameter in the same way as \citet{Nakanishi06}.
As seen in Paper I, VV 219 and VV 254 have a peak of H\emissiontype{I} gas at their overlapping region.
Especially VV 219, the fact that H\emissiontype{I} gas at the contact region is not extended outward the CO disc implies that H\emissiontype{I} gas is accumulated there, while H\emissiontype{I} gas is extended like normal spiral galaxies in the southern disc of NGC 4568.
These features cannot be explained without the interaction between interstellar gas.
Therefore, during a galaxy interaction, interstellar gas would also interact.
Thus, we may expect external pressure in their environments.
The pressure term $P$ with this external pressure can be rewritten instead of the equation (\ref{press}) as follows:
\begin{equation}
	\frac{P}{P_{\odot}}=\left(\frac{\Sigma_{\mathrm{gas}}}{\Sigma_{\mathrm{gas}_{\odot}}}\right)^{2}+\varepsilon^{2},
\end{equation} 
where $\varepsilon^{2}$ is the contribution of the external pressure.
Therefore, in our simulations, free parameters become metallicity $Z$ and external pressure $\varepsilon$.
Figure \ref{fmolexpressure} illustrates a plot of $f_{\mathrm{mol}}$ against the surface density of the total gas with different external pressure $\varepsilon$ assuming metallicity $Z = 1.0 \ Z_{\odot}$ and $\Sigma_{\mathrm{SFR}} = 1.0 \ \Sigma_{\mathrm{SFR_{\odot}}}$ according to our new model. 
When external pressure is included, $f_{\mathrm {mol}}$ is enhanced at the surface density of the total gas where external pressure exceeds the intrinsic pressure as compared with the case without external pressure.

Model fitting procedures were performed as follows:
\begin{quote}
	1. Input the observed $f_{\mathrm{mol}}$, $\Sigma_{\mathrm{gas}}$ and $\Sigma_{\mathrm{SFR}}$ for pixels where CO and H\emissiontype{I} are detected with the S/N ratio higher than 3.\\
	2. For a certain metallicity $Z$ which is fixed throughout a galaxy, calculate $f_{\mathrm{mol}}$ based on our new model using the observed $\Sigma_{\mathrm{gas}}$ and $\Sigma_{\mathrm{SFR}}$\footnote{Each item in the pixel data has the same metallicity in this calculation.}.\\
	3. Subtract theoretical $f_{\mathrm{mol}}$ from the observed $f_{\mathrm{mol}}$ and derive $\chi^{2}$ where CO and H\emissiontype{I} are detected.\\
	4. Change metallicity and external pressure. Both parameters are changed with an interval of 0.1.\\
	5. Iterate above processes within a range of 1 $\sigma$ uncertainty of $Z$ derived from $B$-band luminosity and $\varepsilon$ = 0.0 -- 24.9.\\
	6. Adopt a pair of $Z$ and $\varepsilon$ which achieve the minimum $\chi^{2}$ as the best-fitted parameters.
\end{quote}

The best-fitted parameters are summarised in table \ref{fit}.
The difference between the observed value and theoretical one at one pixel can be obtained substituting the value at the same surface density of the total gas.
In other words, the black dots and the red crosses on any vertical line are derived from the same pixel data.
We will discuss the results of each galaxy in the next section.
\begin{table*}[tbp]
	\centering
	\tbl{A summary of best-fitted parameters}{
			\begin{tabular}{cccclccl}
			\hline
			       	Pair name& Galaxy & \multicolumn{3}{c}{Galactic $X_{\mathrm{CO}}$ factor} & \multicolumn{3}{c}{1/2 $\times$ Galactic $X_{\mathrm{CO}}$ factor}  \\
				       	&	& $Z$ ($Z_{\odot}$) & $\varepsilon$ & minimum $\chi^{2}$ & $Z$ ($Z_{\odot}$) & $\varepsilon$ & minimum $\chi^{2}$\\
			\hline
			Arp 84 & NGC 5394 & 2.5 & 0.0 & 5.3 $\times$ 10$^{-3}$ & 2.5 & 3.4 & 6.2 $\times$ 10$^{-2}$\\
			& NGC 5395 & 1.0 & 2.0 & 2.0 $\times$ 10$^{0}$ & 1.0 & 1.6 & 3.5 $\times$ 10$^{0}$\\
			\hline
			VV 219& NGC 4567 & 1.5 & 6.9 & 4.2 $\times$ 10$^{-1}$ & 1.3 & 0.0 & 9.6 $\times$ 10$^{-1}$\\
			&NGC 4568 & 2.1 & 2.3 & 1.2 $\times$ 10$^{0}$ & 1.2 & 0.0 & 7.1$ \times$ 10$^{0}$\\
			\hline
			VV 254&UGC 12914 & 1.5 & 4.6 & 2.3 $\times$ 10$^{0}$ & 1.3 & 1.1 & 1.5 $\times$ 10$^{-1}$\\
			&UGC 12915 & 1.8 & 0.1 & 1.2$ \times$ 10$^{0}$ & 2.4 & 0.6 & 3.4 $\times$ 10$^{0}$\\
			&The Bridge Region & 0.7 & 4.2 & 2.1 $\times$ 10$^{-1}$ & 0.9 & 2.6 & 5.0 $\times$ 10$^{-1}$\\
			\hline
			The Antennae Galaxies && 1.9 & 10.5 & 9.7 $\times$ 10$^{-2}$ & 2.3 & 11.5 & 2.1 $\times$ 10$^{-1}$\\
			\hline
			\end{tabular}}
			\label{fit}
	
\end{table*}

\subsection{Fitting results of $f_{\mathrm{mol}}$}
\label{fmoldiscuss2}
\begin{figure}[tbp]
	\centering
		\FigureFile(80mm,120mm){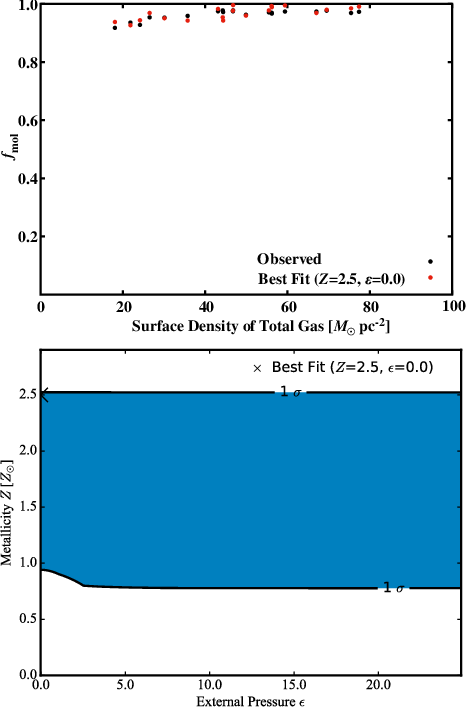}
		\caption{Top: The results of theoretical $f_{\mathrm{mol}}$ fitting for NGC 5394.
						 Free parameters are metallicity $Z$ and external pressure $\varepsilon$. Black dots are observed $f_{\mathrm{mol}}$, while red points are best-fitted theoretical $f_{\mathrm{mol}}$. Bottom: $\chi^{2}$ map of the fitting. The cross represents the best-fitted parameters. The contour is minimum $\chi^{2}$ + 2.3, showing 1 $\sigma$ significance level.}
		\label{ngc5394fmolfit}
	
\end{figure}

\begin{figure}[tbp]
	\centering
		\FigureFile(80mm,120mm){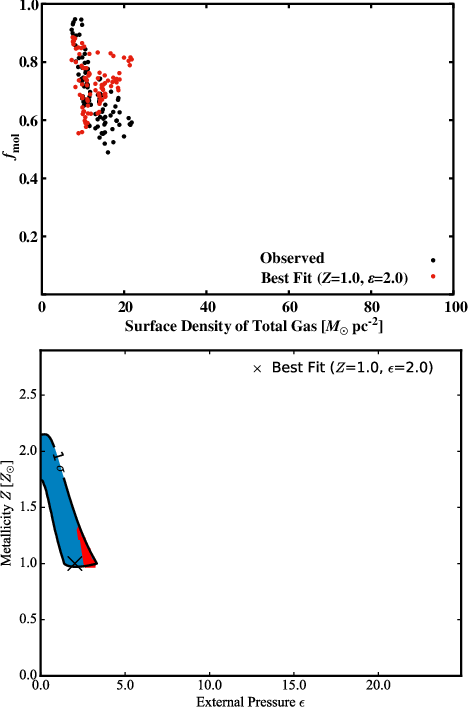}
		\caption{Same as figure \ref{ngc5394fmolfit} but for NGC 5395. The explanation of a red shaded region is described in the text.}
		\label{ngc5395fmolfit}
	
\end{figure}

\begin{figure}[tbp]
	\centering
		\FigureFile(80mm,120mm){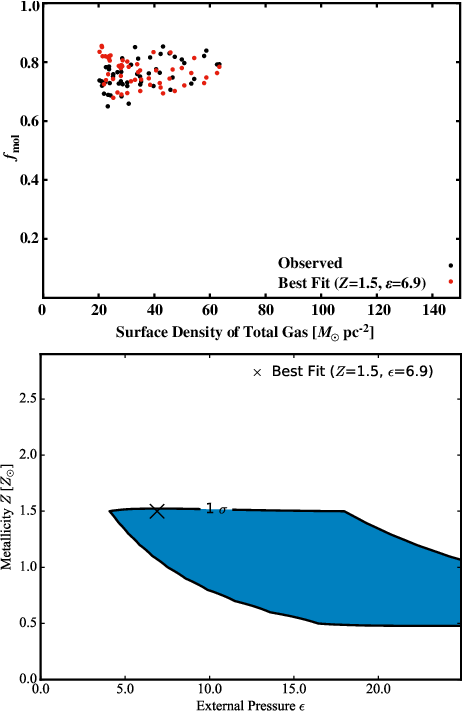}
		\caption{Same as figure \ref{ngc5394fmolfit} but for NGC 4567.}
		\label{ngc4567fmolfit}
	
\end{figure}

\begin{figure}[tbp]
	\centering
		\FigureFile(80mm,120mm){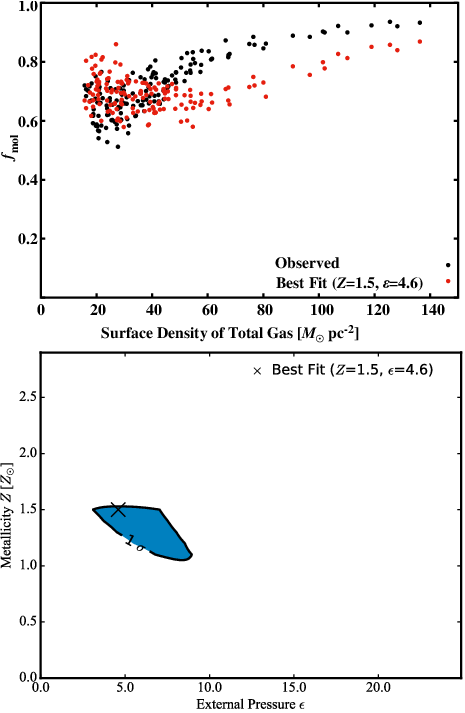}
		\caption{ Same as figure \ref{ngc5394fmolfit} but for NGC 4568.}
		\label{ngc4568fmolfit}
	
\end{figure}

\begin{figure}[tbp]
	\centering
		\FigureFile(80mm,120mm){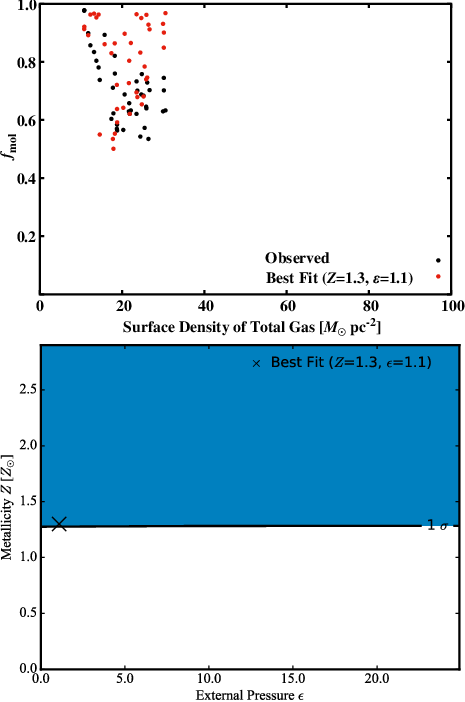}
		\caption{Same as figure \ref{ngc5394fmolfit} but for UGC 12914.}
		\label{ugc12914fmolfit}
	
\end{figure}

\begin{figure}[tbp]
	\centering
		\FigureFile(80mm,120mm){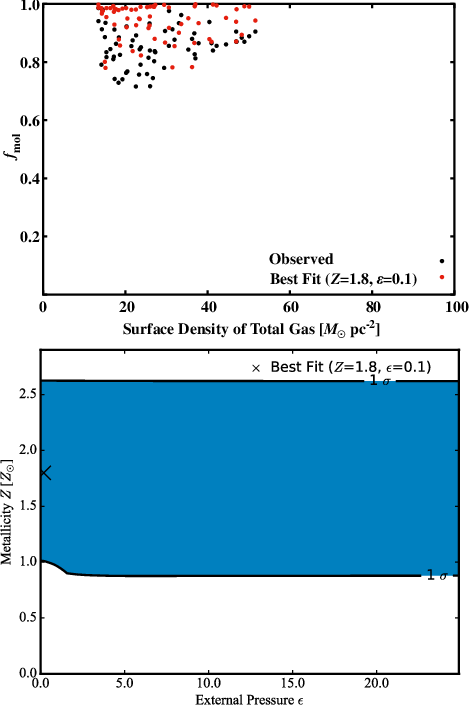}
		\caption{ Same as figure \ref{ngc5394fmolfit} but for UGC 12915.}
		\label{ugc12915fmolfit}
	
\end{figure}

\begin{figure}[tbp]
	\centering
		\FigureFile(80mm,120mm){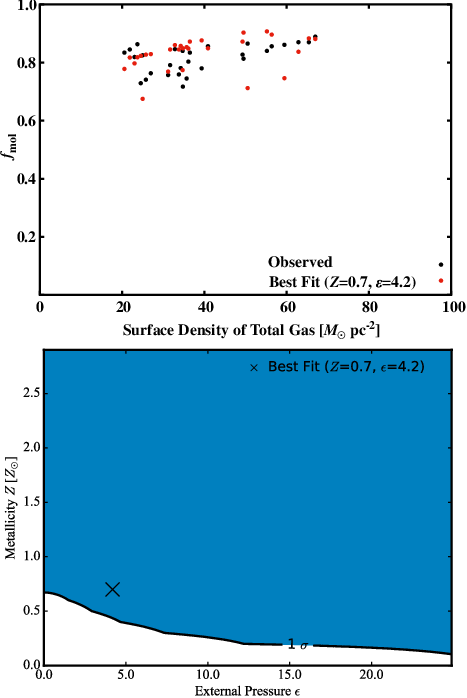}
		\caption{Same as figure \ref{ngc5394fmolfit} but for the overlapping region of VV 254. Metallicity data of this region is not available.}
		\label{overlapfmolfit}
	
\end{figure}

\begin{figure}[tbp]
	\centering
		\FigureFile(80mm,120mm){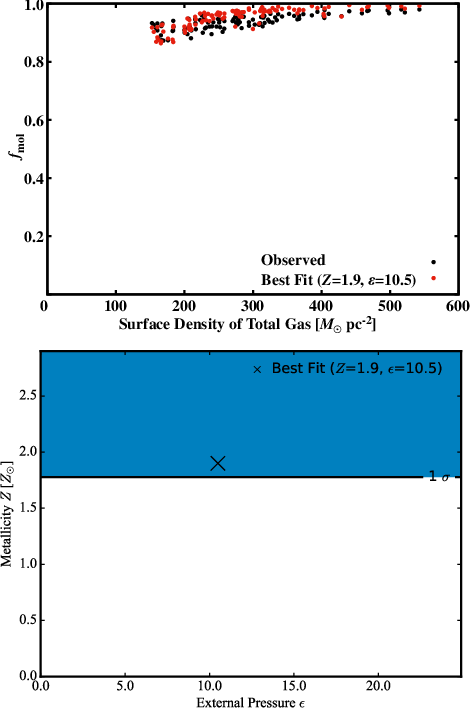}
		\caption{ Same as figure \ref{ngc5394fmolfit} but for the Antennae Galaxies.}
		\label{antennaefmolfit}
	
\end{figure}

Figures \ref{ngc5394fmolfit}--\ref{antennaefmolfit} show the results of the model fitting for each galaxy.
The bottom of each figure represents the $\chi^{2}$ map.
The blue region enclosed by a contour whose value is the minimum $\chi^{2}$ + 2.30 indicates 1 $\sigma$ significance level of the fitting.
NGC 5395 and NGC 4568 are constrained within a narrow range of parameters.
Although $f_{\mathrm{mol}}$--$\Sigma_{\mathrm {gas}}$ relation for NGC 4567 can be explained by a wide range of parameter sets, external pressure is necessary to reproduce the observed relation.
Other galaxies (NGC 4567, UGC 12914, UGC 12915, the overlapping region of VV 254, and the Antennae Galaxies) could not impose any limitation on parameters,
and in particular, the external pressure term.
We will see the meaning of the fitting results in the successful cases first, and next, discuss the reasons for failure where the fitting does not work well.

As mentioned in section \ref{localfmol}, the monotonic decrease of $f_{\mathrm{mol}}$ with the increase of the surface density of the total gas is seen in NGC 5395.
However, plausible parameter sets shown in figure \ref{ngc5395fmolfit} may contain parameters which cannot reproduce the decreasing trend of $f_{\mathrm{mol}}$ along with an increase of surface density of the total gas.
In fact, the parameter set of ($Z/Z_{\odot}$, $\varepsilon$)= (1.9, 0.0) included in the plausible parameters unsuccessfully reproduce the trend as we conclude in sub-section \ref{fmoldiscuss} (see, figure \ref{ngc5395woExt}).
In order to check which parameter set can reproduce the observed trend, we quantify the simulated parameter set.
Since the range of the surface density of total gas of NGC 5395 is about from 5  $\MO$ pc$^{-2}$ to 25 $\MO$ pc$^{-2}$, we divided data points into three parts: the data with 10.0 $> \Sigma_{\mathrm{gas}}$ (low $\Sigma_{\mathrm{gas}}$), 15.0 $> \Sigma_{\mathrm{gas}} >$ 10.0 (middle $\Sigma_{\mathrm{gas}}$), and $\Sigma_{\mathrm{gas}} >$15.0 (high $\Sigma_{\mathrm{gas}}$). 
Then, theoretically derived $f_{\mathrm{mol}}$ are averaged over each part ($f_{\mathrm{mol}}^{\mathrm{ave}}$).
We checked for each parameter set whether theoretical $f_{\mathrm{mol}}$ monotonically decreases along with the increase of $\Sigma_{\mathrm{gas}}$, i.e., check whether the parameter sets fulfil 
\begin{equation}
	f_{\mathrm{mol}}^{\mathrm{ave}}(\mathrm{low} \ \Sigma_{\mathrm{gas}}) > f_{\mathrm{mol}}^{\mathrm{ave}}(\mathrm{middle} \ \Sigma_{\mathrm{gas}}) > f_{\mathrm{mol}}^{\mathrm{ave}}(\mathrm{high} \ \Sigma_{\mathrm{gas}}).
\end{equation}
The parameter sets that show the trend of the monotonous decrease of $f_{\mathrm{mol}}$ along with the increase of $\Sigma_{\mathrm{gas}}$ are shown in figure \ref{ngc5395fmolfit} bottom with a red shade.
This figure illustrates how our model can reproduce the odd trends in NGC 5395 in very narrow parameter space and no parameter set can reproduce the observed trend if the external pressure term is not included.

The fact that the external pressure is needed to reproduce the observed trend of NGC 5395 can be qualitatively understood as follows.
In the case of constant SFR (see, figure \ref{fmolmetallicity}), 
$f_{\mathrm{mol}}$ increases gradually along with the increase in the surface density of total gas.
In reality, SFR is not constant throughout a galaxy.
When SFR in NGC 5395 becomes high along with the increase in the surface density of total gas, as is often the case with normal spiral galaxies (e.g., \cite{Kennicutt87}), 
$f_{\mathrm{mol}}$ at a high surface density of total gas should get lower than that in the case of constant SFR.
Additionally, the increase in external pressure on $f_{\mathrm{mol}}$ selectively influences where the surface density of total gas is low (figure \ref{fmolexpressure}).
Due to the combination of these effects, NGC 5395 has an odd trend of the decrease in $f_{\mathrm{mol}}$ with the increase in the surface density.
This result also suggests that the trend of decreasing with the surface density of total gas in NGC 5395 is not due to the uncertainty of the inclination of molecular gas and atomic gas discs, and there is an influence of the external pressure. 
The $\chi^{2}$ map (figure \ref{ngc5395fmolfit} bottom) suggests possible parameters are located in a very narrow range.
Metallicity equal to or slightly higher than the solar metallicity with external pressure (4.0 $> \varepsilon >$ 2.0) is necessary to explain the odd tendency of NGC 5395.

Plausible parameter space of both NGC 4567 and NGC 4568 shows external pressure higher than 2.0, corresponding to 16.0 $\MO$ pc$^{-2}$ (figure \ref{ngc4567fmolfit} and \ref{ngc4568fmolfit}).
Although the NGC 4567 and NGC 4568 pair has the most undisturbed morphology in our sample, the fitting results demand external pressure.
These results suggest that a close encounter may cause external pressure even in early stages of interaction.

NGC 5394, the VV 254 system, and the Antennae Galaxies cannot restrict the parameter space.
Therefore, it is not clear whether these galaxies are affected by external pressure.
There are several reasons to explain the model fitting failure.

High $f_{\mathrm{mol}}$ value nearly 1.0 (fully molecules) throughout the region where CO and H\emissiontype{I} are detected (see, H$\alpha$, FUV, 24 $\mu$m and 8 $\mu$m images in Paper I) is observed in NGC 5394 and the Antennae Galaxies. 
Since NGC 5394 and the Antennae Galaxies show active star formation even in early or mid stages of interaction,
these galaxies experience strong UV photodissociation.
According to our model, it is difficult to distinguish the cause of high $f_{\mathrm{mol}}$ from high metallicity and high external pressure as seen in figure \ref{fmolmetallicity} and figure \ref{fmolexpressure}.
This is the reason the model fitting could not constrain parameters in these galaxies.

VV 254 system can be explained in other ways.
Elmegreen's model assumes that H\emissiontype{I}--H$_{2}$ production and destruction rate is under equilibrium.
VV 254 undertakes the first encounter just 2 $\times$ 10$^{7}$ years ago \citep{Condon93}.
Gravitational disturbance changes the environment of galaxies.
Although it is enough to produce molecular gas from interstellar gas (this issue will be discussed in section \ref{mechanismfmol}), it may not reach the balanced state of H\emissiontype{I}--H$_{2}$ production and destruction rate. 
Thus, the facts that it is observed large $f_{\mathrm{mol}}$ dispersion at a same surface density of total gas and unsuccess in the model fitting could be reasonable for VV 254.
The second possibility is the uncertainty of SFR.
According to \citet{Kennicutt98}, UV emission, which we used to derive SFR for this pair, traces star formation over time-scales of 10$^{8}$ years or shorter.
Therefore, SFR derived from UV emission traces star formation not only occurred after the collision but also before the collision.
Thus SFR that we used for the fitting should not trace recent UV radiation.
Moreover, recent Pa$\alpha$ observations imply that star formation in VV 254 is more active than our estimation \citep{Komugi12}.
The fitting results for UGC 12915 (figure \ref{ugc12915fmolfit}) showed that 
calculated $f_{\mathrm{mol}}$ tends to have a higher value than observationally derived $f_{\mathrm{mol}}$.
This could be due to an underestimation of SFR through FUV emission.


\subsection{The effects of different $X_{\mathrm{CO}}$ factor}
\label{Xco}
We focus on the uncertainty of the conversion factor, $X_{\mathrm{CO}}$, from $I_{\mathrm{CO}}$ to $N_{\mathrm{H_{2}}}$.
Although we assumed the same conversion factor for interacting galaxies and isolated galaxies, observations and simulations report that interacting galaxies have lower conversion factor compared to normal spiral galaxies \citep{Zhu03,Zhu07,Narayanan12}.
Thus, there is a possibility that we overestimate molecular gas mass and $f_{\mathrm{mol}}^{\mathrm{global}}$ for interacting galaxies.
If we assume that the conversion factor is a factor of 2 lower than the conventional value we adopted in this discussion, that is, $X_{\mathrm{CO}}$ = 9.0 $\times$ 10$^{19}$ cm$^{-2}$ (K km s$^{-1}$)$^{-1}$,
we find that the average of $f_{\mathrm{mol}}^{\mathrm{global}}$ is 0.62 $\pm$ 0.17 and the median is 0.59.
In a case of assuming a factor of 2 lower $X_{\mathrm{CO}}$, $f_{\mathrm{mol}}^{\mathrm{global}}$ of interacting galaxies is as same as that of the control sample galaxies (0.52 $\pm$ 0.18).

Then we performed the theoretical fitting on $f_{\mathrm{mol}}$--$\Sigma_{\mathrm{gas}}$ relation again in the case of adopting $X_{\mathrm{CO}}$ of 9.0 $\times$ 10$^{19}$ cm$^{-2}$ (K km s$^{-1}$)$^{-1}$ in order to check whether theoretical model fitting supports the standard $X_{\mathrm{CO}}$ conversion factor.
The fitting results show that a factor of 2 lower $X_{\mathrm{CO}}$ yields twice worse the minimum $\chi^{2}$ than the standard $X_{\mathrm{CO}}$ factor for all interacting galaxies on average (table \ref{fit}).
Large velocity gradient (LVG) analyses \citep{Zhu03,Zhu07} show that VV 254 and the Antennae Galaxies have lower $X_{\mathrm{CO}}$ than the Galactic one.
Contrary to the result of $f_{\mathrm{mol}}^{\mathrm{global}}$ and previous LVG analyses, the theoretical fitting suggests that it may be better to adopt the Galactic $I_{\mathrm{CO}}$ to $N_{\mathrm{H_{2}}}$ conversion factor in interacting galaxies in early stages rather than a smaller $X_{\mathrm{CO}}$ conversion factor.
However, since the conversion factor in interacting galaxies is still under discussion, more research is needed to confirm our assumption.

\subsection{Mechanism of enhancement of $f_{\mathrm{mol}}$ in interacting galaxies}
\label{mechanismfmol}
We discuss possible mechanisms of the enhancement of $f_{\mathrm{mol}}$ seen in interacting galaxies even in the early stage in terms of a shock-induced molecular gas formation.
One possible scenario is that widespread shock in the galaxies compresses the interstellar gas. 
In order to understand widespread star formation throughout interacting galaxies as seen in the Antennae Galaxies
 which cannot be explained by gas concentration models (e.g., \cite{BH96}),
\citet{Barnes04} made a star-formation model in which star formation takes place where the rate of shock-induced mechanical heating changes substantially.
His model successfully explained the widespread star formation seen in the Mice galaxy (NGC 4676) which is in the early stage of the interaction is due to the shock-induced star formation, and it lasts about 10$^{8}$ years. 
Additionally, \citet{Icke85} shows by numerical hydrodynamic calculations that shocks occur in a disc of gas in a gravitationally perturbed galaxy at the early stage of the interaction even in a distant encounter like fly-by interactions.
In Icke's simulations, strong shock emerges even in the minor merger whose mass ratio is 4.
The fact that it is observed a shock feature in secondary galaxy NGC 5394 which is one of our samples \citep{Roche15} reinforces these simulations.
In the result, it can be thought that high pressure is brought to the galaxy.
This pressure induced by the interaction with the counterpart galaxy should enhance the conversion rate from H\emissiontype{I} to H$_{2}$, and 
high $f_{\mathrm{mol}}$ is achieved even in the low surface density of total gas.

Another scenario is a cloud-cloud collision model that should suit the contact region of VV 219 and the overlapping region of VV 254.
This cloud-cloud collision model has two cases of the molecular gas formation.
One difference between the two is behaviours of giant molecular clouds (GMCs) during the galaxy-galaxy collision.
\citet{JS92} investigate the behaviours of interstellar clouds during a galaxy interaction and 
show that only H\emissiontype{I} clouds can collide with each other and GMCs cannot collide because the volume-filling factor is smaller than H\emissiontype{I} clouds.
The collision between H\emissiontype{I} clouds causes strong shock and clouds fully ionised.
Since such an ionised gas compresses a surface of un-collided GMCs with their high pressure and GMCs become unstable, 
they attribute effective star formation in interacting galaxies to this mechanism.
Although they do not consider the reaction between high pressure and pre-collision H\emissiontype{I} clouds, this leads to the production of molecular gas from the H\emissiontype{I} clouds.

On the other hand, \citet{Braine04} considered the possibility of GMC-GMC collisions and 
concluded that not only H\emissiontype{I} clouds but GMCs can also collide because the volume-filling factor which \citet{JS92} assumed is too small.
When GMC-GMC collisions happen, GMCs are destroyed and ionised with local shock as H\emissiontype{I} clouds rather than encouraged to form stars and then $f_{\mathrm{mol}}$ decreases.
Hot ionised gas with temperature of $\simeq $ 10$^{6}$ K, however, becomes rapidly below 10$^{4}$ K through emission line \citep{Harwit87}.
The cooling time-scale is only a few years and much shorter than the collision time-scale.
Therefore, ionised gas, whether originally atomic or molecular gas, soon becomes cool and neutral.

In these scenarios, the key process produced by the interaction is shock. 
If the reproduced H\emissiontype{I} gas is dense enough, H\emissiontype{I} becomes H$_{2}$.
H$_{2}$ formation time-scale is governed by the inverse of the density. 
The typical time-scale for H\emissiontype{I}-H$_{2}$ transition in diffuse ISM after shock is not longer than an order of 10$^{7}$ yr \citep{Bergin04}.
The density in the collision front including the overlapping region is expected to be higher than the disc region of the progenitors because that region is produced in the result of the collision of two galaxies. 
This suggests that the time-scale for H\emissiontype{I}-H$_{2}$ transition by the shock during an interaction event is shorter than typical one.
This time-scale is comparable or shorter than the collision time-scale of one encounter ($\sim$ 10$^{7}$ yr).
We can regard that the pre-exist H\emissiontype{I} and H$_{2}$ clouds co-exist during the encounter.
Thus, no matter which HI or H$_{2}$ gas before the collision, the efficient production of molecular gas procedure occurs in interacting galaxies.

However, in the case of GMC-GMC collisions, CO gas that we observed and used as an H$_{2}$ tracer may also be destroyed. 
If it is the case, a production time-scale of H$_{2}$ may differ from that of typical one, which assumes that CO already exists.
Here, we discuss the time-scale of molecular gas production based on the chemical processes under the condition that molecular gas is destroyed by a GMC-GMC collision.
In this case, since both H$_{2}$ and CO gas are destroyed, we must consider time-scales of the production of H$_{2}$ and CO gas.
If the time-scale of the formation of CO greatly differs from that of H$_{2}$, the CO-H$_{2}$ conversion factor is changed, and it leads to a mistake in estimating real $f_{\mathrm{mol}}$.
Although there are many ways to convert from C\emissiontype{I} to CO, the transition we must pay attention to is the dominant transition on the production of CO.
Thus we treat one of the most contributive routes on the production of CO gas that occurs via CH$_{2}^{+}$. 
This process progresses in the photodissociation region and CH$_{2}^{+}$ is made from H$_{2}$ and C$^{+}$. 
According to this process, the time-scale of the formation of CO gas is $\sim$ 10$^{8} \ n_{\mathrm{H}}^{-1}$ years, 
where $n_{\mathrm{H}}$ is a number density of hydrogen in the unit of cm$^{-3}$ \citep{TH85, Suzuki92,Oka04},
while the time-scale of the formation of H$_{2}$ gas is $\sim$ 10$^{9} \ n_{\mathrm{H}}^{-1}$ years. 
Thus in typical molecular cloud whose density is $\sim$10$^{2}$ cm$^{-3}$, these time-scales correspond to $\sim$ 10$^{6}$ years and $\sim$ 10$^{7}$ years for CO and H$_{2}$ gas, respectively.
Since H$_{2}$ regulates the CO formation in this process, CO gas can be made only after the production of H$_{2}$ gas.
Thereby, the order of time-scales to form CO and H$_{2}$ both take about 10$^{7}$ years.
Therefore, we do not underestimate the CO-H$_{2}$ conversion factor at least in later than 10$^{7}$ years after the collision.
Even for VV 254 which is just after the collision, the time-scale for the production of CO and H$_{2}$ of 10$^{7}$ years is comparable to the collision age. 
Therefore, high $f_{\mathrm{mol}}$ in the interacting galaxies is not the result of a misunderstanding due to the difference of the CO-H$_{2}$ conversion factor but for a real tendency.
Moreover, a shock induced by the collision is capable of shortening the time-scale of the formation of molecular gas.
\citet{Guillard09} discussed the time-scale to form H$_{2}$ gas under strong shock 
to explain warm H$_{2}$ rotation line emission found in the collision front of ongoing interacting galaxies, the Stephan's Quintet.
According to them, the time-scale of the production of H$_{2}$ from hot plasma is roughly the inverse of the gas density with temperature of 10$^{4}$ K and is formulated as follows:
\begin{equation}
	t_{\mathrm{H_{2}}} \mathrm{[yr]} \simeq 7 \times 10^{5} f_{\mathrm{dust}}\left(\frac{2 \times 10^{5} \mathrm{[K \ cm^{-3}]}}{P_{\mathrm{th}}}\right)^{0.95},
\end{equation}
where $P_{\mathrm{th}}$ is the thermal gas pressure and $f_{\mathrm{dust}}$ the dust mass fraction remaining in the gas.
They showed that warm H$_{2}$ gas is formed in the collision front of the Stephan's Quintet even at its age after the collision of 5 $\times$ 10$^{5}$ years.
Furthermore, time-scales for the formation of CO and cold H$_{2}$ take only $\sim$ 10$^{5}$ years after temperature of the postshock gas falls below 10$^{4}$ K (see, \citet{Guillard09} figure C.2).
Taking this fact into account, it is not surprising that plenty of cold H$_{2}$ gas and CO gas is made in interacting galaxies with the age of the interaction being
much longer than Stephan's Quintet.
In our observational results, the overlapping region of VV 254 would be the case that can be explainable with this idea.
Although strong shock should prevail there because VV 254 undergoes a head-on collision, 
molecular gas can be formed from ionised gas in a short time, something that is also briefly mentioned in \citet{Braine04}.
Besides, $f_{\mathrm{mol}}$ should vary with the location 
because the time-scale of the formation of molecular gas depends upon a pre-shock gas density and time after the collision also differs according to the location.
Thus, the dispersion of $f_{\mathrm{mol}}$ at the same surface density of total gas is expected to become large in interacting galaxies under the existence of the shock.
From the perspective of chemical time-scale, it is not ruled out that high $f_{\mathrm{mol}}$ seen in interacting galaxies is achieved by the shock induced by the interaction.
Therefore, even if GMC-GMC collisions happen, $f_{\mathrm{mol}}$ may increase and 
when the density is sufficiently high $f_{\mathrm{mol}}$ is seen even in the early stage of the interaction.

We have discussed why external pressure in interacting galaxies in early stages increases.
Two scenarios, widespread shock and cloud-cloud collisions (both H\emissiontype{I}-H\emissiontype{I} cloud collisions and GMC-GMC collisions), are plausible causes of external pressure.
The former one explains galactic scale external pressure, while the latter is particularly suits for the collision front of the interaction.
In both cases, shock plays an important role in effective gas phase transition from H\emissiontype{I} to H$_{2}$.

\section{Summary}
We investigated the ISM properties in interacting galaxies in early and mid stage of the interaction using CO and H\emissiontype{I} mapping data.
Our findings are as follows:\\\\
1. Molecular hydrogen gas and total gas mass normalised with \textit{Ks}-band luminosity are unchanged between the interacting galaxies of our sample and isolated galaxies. 
This is due to ionisation of atomic gas and/or production of molecular gas from atomic gas.\\
2. The global molecular gas fraction in the interacting galaxies of our sample is higher than that in field galaxies.\\
3. Few interacting galaxies show the radial decrease in local molecular gas fraction seen in typical isolated galaxies.\\
4. A pixel-to-pixel comparison showed that high molecular gas fraction is achieved even if the surface density of total gas is low.\\
5. The local molecular gas fraction decreases along with the increase of the surface density of total gas is found in NGC 5395.
This tendency has never been observed.\\
6. With numerical model fittings using observable parameters, NGC 5395 and NGC 4568 are successfully explained the observation, which may be caused by external pressure.\\
7. Adopting a half Galactic CO-H$_{2}$ conversion factor instead of the Galactic one, results of numerical model fitting for all target interacting galaxies get worse, implying that CO-H$_{2}$ conversion factor in the interacting galaxies in the early stage is unchanged.\\
8. The cause of external pressure can be explained by widespread shocks on galactic scale and by cloud-cloud collisions in colliding regions. Shocks plays an important role in both scenarios.\\
9. Chemical time-scale for the production of both CO and H$_{2}$ does not reject an increase of molecular gas fraction even if cloud-cloud collisions destroy pre-existing molecular clouds.

\end{document}